\documentclass[twocolumn]{aastex631}

\usepackage{CJK}

\shorttitle{Transmission spectrum of HAT-P-1b}
\shortauthors{Chen et al.}
\graphicspath{{./}{figures/}}
\begin{document}
\begin{CJK*}{UTF8}{gbsn}

\title{Detection of Na and K in the atmosphere of the hot Jupiter HAT-P-1b with P200/DBSP}

\correspondingauthor{Guo Chen}
\email{guochen@pmo.ac.cn}

\author[0000-0003-0740-5433]{Guo Chen (陈果)}
\affiliation{CAS Key Laboratory of Planetary Sciences, Purple Mountain Observatory, Chinese Academy of Sciences, Nanjing 210023, People's Republic of China}
\affiliation{CAS Center for Excellence in Comparative Planetology, Hefei 230026, People's Republic of China}

\author[0000-0003-0746-7968]{Hongchi Wang (王红池)}
\affiliation{CAS Key Laboratory of Radio Astronomy, Purple Mountain Observatory, Chinese Academy of Sciences, Nanjing 210023, People's Republic of China}

\author[0000-0002-2190-3108]{Roy van Boekel}
\affiliation{Max-Planck-Institut f\"{u}r Astronomie, K\"{o}nigstuhl 17, 69117 Heidelberg, Germany}

\author[0000-0003-0987-1593]{Enric Pall\'{e}}
\affiliation{Instituto de Astrof\'{i}sica de Canarias, V\'{i}a L\'{a}ctea s/n, E-38205 La Laguna, Tenerife, Spain}
\affiliation{Departamento de Astrof\'{i}sica, Universidad de La Laguna, E-38206 La Laguna, Tenerife, Spain}



\begin{abstract}

We present a new optical transmission spectrum of the hot Jupiter HAT-P-1b based on two transits observed with the Double Spectrograph (DBSP) on the Palomar 200-inch (P200) telescope. The DBSP transmission spectrum, covering a wavelength range from 3250 to 10007 \AA, is consistent with that observed with the Hubble Space Telescope (HST), but the former has a finer spectral resolution. The DBSP spectrum alone reveals the presence of a pressure broadened line wing for Na, the line cores for both Na and K, and tentative evidence for H$_2$O. We obtain consistent results from the spectral retrieval analyses performed on the DBSP-only dataset and the DBSP, HST, and Spitzer combined dataset. Our retrievals suggest a mostly clear atmosphere for HAT-P-1b, with a cloud coverage of $22^{+5}_{-3}$\% that is dominated by enhanced haze. We derive subsolar abundances for Na, K, and C, and subsolar-to-solar for O. Future observations with James Webb Space Telescope and ground-based high-resolution spectrographs should be able to not only confirm the presence of these species but also stringently constrain the formation and migration pathways for HAT-P-1b.

\end{abstract}

%\keywords{Exoplanet atmospheres --- Exoplanet atmospheric composition --- Transmission spectroscopy --- Hot Jupiters --- Exoplanets}

\keywords{\href{http://astrothesaurus.org/uat/487}{Exoplanet atmospheres (487)}; 
\href{http://astrothesaurus.org/uat/2021}{Exoplanet atmospheric composition (2021)}; 
\href{http://astrothesaurus.org/uat/2133}{Transmission spectroscopy (2133)}; 
\href{http://astrothesaurus.org/uat/753}{Hot Jupiters (753)}; 
\href{http://astrothesaurus.org/uat/498}{Exoplanets (498)}
}


%------------------------------------------------------------------------
%  1. INTRODUCTION
%------------------------------------------------------------------------
\section{Introduction}
\label{sec:intro}
\end{CJK*}

Observational characterization of exoplanet atmospheres has long been pursued since the first discovery of non-transiting and transiting exoplanets around main-sequence stars \citep{1995Natur.378..355M,2000ApJ...529L..45C,2002ApJ...568..377C,2003Natur.422..143V}. With the continuous developments of transit spectrophotometry, direct spectroscopy, and high-resolution Doppler spectroscopy, abundant spectral features induced by atoms and molecules have been detected in dozens of exoplanets \citep{2019ARA&A..57..617M}. The ultrahot gas giants are generally found to be dominated by ionized or neutral atoms with inverted thermal profiles, while the cooler ones by molecules without inversions \citep{2022A&A...662A.101S,2022ApJS..260....3C}. 

Among the various species detected in exoplanet atmospheres, Na and K are important to understand rainout process \citep{2017ApJ...848...83L,2022arXiv220413714M}, photoionization \citep{2014ApJ...796...15L}, radiative cooling \citep{2017ApJ...851..150H}, and wind patterns \citep{2020A&A...633A..86S,2021A&A...653A..73S}. Meanwhile, Na and K play a crucial role in constraining the reference pressure levels \citep{2019AJ....157..206W,2019MNRAS.482.1485P} and cloud properties \citep{2017ApJ...849...72K,2021AJ....162..179M} with their pressure broadened line wings. The inference of pressure broadening is still rare \citep{2018Natur.557..526N,2018A&A...616A.145C,2019AJ....157...21P,2020A&A...642A..54C,2021ApJ...906L..10A,2022MNRAS.510.4857A}, which relies on high-quality transit spectrophotometry in the optical. 

HAT-P-1b is a classical hot Jupiter orbiting the secondary component of a G type binary system \citep{2007ApJ...656..552B}. With an inflated radius of 1.319~R$_\mathrm{Jup}$, a relatively low mass of 0.525~M$_\mathrm{Jup}$, and an equilibrium temperature of 1322~K \citep{2014MNRAS.437...46N}, HAT-P-1b is expected to have an atmospheric scale height of $H/R_\star=0.00078$ that can produce large amplitudes of transmission spectral features in clear atmospheres. Located in a parameter space similar to the hot Jupiters exhibiting pressure broadening \citep[i.e., 1200--1500~K, 0.2--0.6~M$_\mathrm{Jup}$, $>$1.0~R$_\mathrm{Jup}$;][]{2020A&A...642A..54C}, HAT-P-1b has a high priority for the search of Na and K. 

Indeed, several photometric and spectrophotometric campaigns have been conducted to characterize HAT-P-1b's atmospheric compositions \citep{2013MNRAS.435.3481W,2014MNRAS.437...46N,2015MNRAS.450..192W,2015ApJ...811...55M,2016Natur.529...59S,2019AA...631A.169T} and thermal structure \citep{2010ApJ...708..498T,2011A&A...528A..49D}. The transmission spectra acquired with the Hubble space telescope (HST) show the presence of both H$_2$O and Na \citep{2013MNRAS.435.3481W,2014MNRAS.437...46N,2016Natur.529...59S}. In particular, \citet{2014MNRAS.437...46N} reported an excess absorption at Na (3.3$\sigma$) by comparing the 30~\AA\ Na narrow band to the blue and red neighboring 600~\AA\ broad bands. But it is not confirmed in the ground-based observations \citep{2015ApJ...811...55M,2019AA...631A.169T} due to limited spectrophotometric precision. On the other hand, using the 12~\AA\ tunable filters, \citet{2015MNRAS.450..192W} reported a strong detection of K (4.3$\sigma$ at line wing and 6.1$\sigma$ at line core) by comparing the line core and wing to two narrow passbands far outside the K feature. However, such comparison is based on data taken on different dates with distinct systematics and limited out-of-transit measurements. 

In this paper, we report the detection of Na and K in the newly acquired optical transmission spectrum of HAT-P-1b, based on uniformly spaced narrow passbands around the Na and K wavelengths. We present the observation summary in Section \ref{sec:obs} and detail the data reduction in Section \ref{sec:datared}. We describe the light-curve modeling process and derive the transmission spectra in Section \ref{sec:lc}. We perform Bayesian spectral retrieval analyses on the transmission spectra in Section \ref{sec:retrieval}. We compare our results with literature and interpret HAT-P-1b's atmosphere in Section \ref{sec:discuss}. Finally, we present our conclusions in Section \ref{sec:concl}.

%++++++++++++++++++++++++++++++++++
%   Table
%++++++++++++++++++++++++++++++++++
\begin{deluxetable*}{cccccccccc}
\tablecaption{Observation summary\label{tab:obsum}}
\tablehead{
\colhead{Transit} & 
\colhead{Date} & 
\colhead{Start} & 
\colhead{End} & 
\colhead{Dichroic} & 
\colhead{Channel} & 
\colhead{$t_\mathrm{exp}$} & 
\colhead{Airmass} & 
\colhead{Seeing-limited} & 
\colhead{Weather} 
\\
\colhead{\#} &
\colhead{UT} &
\colhead{UT} &
\colhead{UT} &
\colhead{} &
\colhead{} &
\colhead{(sec)} &
\colhead{} &
\colhead{$\Delta\lambda$ at $\lambda$~(\AA)} &
\colhead{}
} 

\startdata
1 & 2012-09-15 & 04:29 & 09:09 &  D52 & Blue & 60 & 1.18-1.00-1.10 & 5.9 (4518) & mostly clear\\
1 & 2012-09-15 & 04:29 & 09:09 &  D52 & Red  & 60 & 1.18-1.00-1.10 & 8.1 (7699) & mostly clear\\
2 & 2016-07-27 & 05:09 & 10:17 &  D68 & Blue & 30 & 2.12-1.01-1.01 & 3.8 (5893) & mostly clear\\
2 & 2016-07-27 & 05:09 & 10:19 &  D68 & Red  & 180 & 2.12-1.01-1.01 & 3.6 (7699) & mostly clear
\enddata
\end{deluxetable*}

%++++++++++++++++++++++++++++++++++
%   Figure
%++++++++++++++++++++++++++++++++++
\begin{figure*}
\centering
\includegraphics[width=0.9\textwidth]{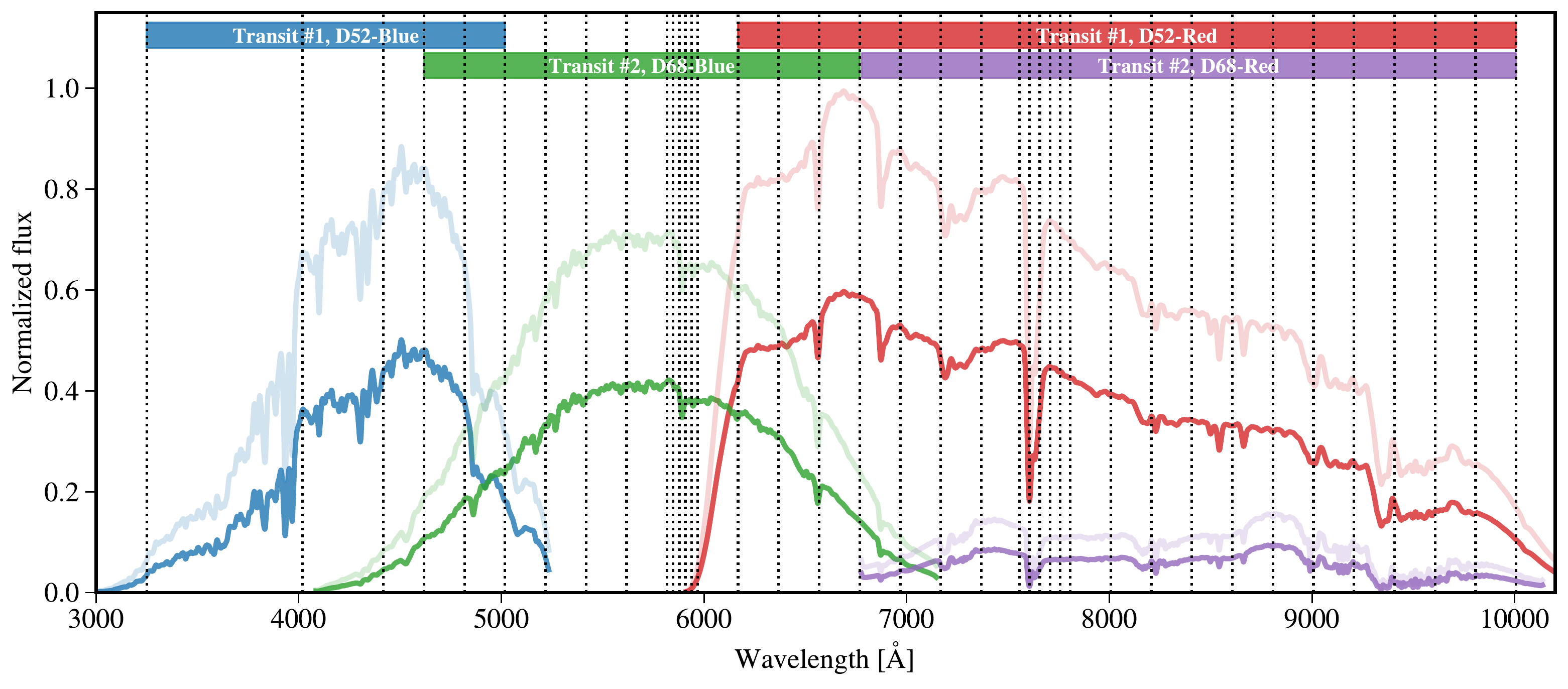}
\caption{Example stellar spectra of HAT-P-1 (dark colored) and its reference star (light colored). Two transits were observed by P200/DBSP with the D52 and D68 dichroics, respectively. Vertical dotted lines indicate the borders of adopted spectroscopic passbands. The blue and red regions show the adopted wavelength ranges to calculate the white light curves for Transit 1, while green and purple for Transit 2, respectively. The spectral resolution has been degraded for display purpose.}
\label{fig:dbsp_spec}
\end{figure*}

%------------------------------------------------------------------------
%  2. OBSERVATIONS
%------------------------------------------------------------------------
\section{Observations}
\label{sec:obs}

We observed two transits of HAT-P-1b using the Double Spectrograph \citep[DBSP;][]{1982PASP...94..586O} installed on the Palomar 200-inch (P200) Hale telescope in California, USA. The target star HAT-P-1 \citep[BD+37 4734B, $G\mathrm{mag}=10.2$;][]{2021A&A...649A...1G} and a reference star \citep[BD+37 4734A, $G\mathrm{mag}=9.6$, 11$\farcs$3 from HAT-P-1;][]{2021A&A...649A...1G} were simultaneously monitored by a 128$''$-long and 10$''$-wide slit. After the slit, the light was splitted into the blue and red channels with a dichroic, and then dispersed by two gratings and recorded by two 2048$\times$4096 CCDs separately. The blue and red CCDs have a pixel scale of 0$\farcs$389 and 0$\farcs$293, respectively. A summary of the two transit observations is given in Table~\ref{tab:obsum}.

The first transit was observed on the night of 2012 September 14 (Transit 1). The dichroic D52 was used to split light at 5200~\AA, along with the 600/4000 (i.e., 600 lines~mm$^{-1}$ blazed at 4000~\AA) grating for the blue channel and the 316/7500 grating for the red channel, resulting in an average dispersion of 1.1~\AA\ (blue) and 1.5~\AA\ (red) per pixel. The order-blocking filter RG610 was employed in the red channel. The observations were conducted continuously for 4.7 hours, covering the complete transit event and $\sim$50 minutes for both pre-ingress and post-egress out-of-transit. The weather condition was mostly clear, but the seeing was moderately variable during the ingress. We measured a range of 1$\farcs$3--2$\farcs$7 and a median value of 1$\farcs$6 at around 7500~\AA\ for the FWHM of the stellar spatial profiles, indicating a seeing-limited spectral resolution of 6.8--13.7~\AA. We adopted an exposure time of 60 seconds for both blue and red channels, and collected a total of 205 (blue) and 201 (red) frames. The spectra of FeAr and HeNeAr arc lamps were acquired through a 1$''$-wide slit for subsequent wavelength calibrations.

The second transit was observed on the night of 2016 July 26 (Transit 2). The dichroic D68 was used. The instrument setup team inadvertently swapped the requested blue and red gratings. Consequently, the observations were conducted with 600/10000 for the blue and 600/4000 for the red, resulting in an average dispersion of 1.1~\AA\ (blue) and 0.8~\AA\ (red) per pixel. The incorrect placement of gratings strongly reduced the photon count rate in the red channel (see Figure~\ref{fig:dbsp_spec}). Both channels were left clear, without any order-blocking filter. The observations lasted for 5.2 hours, covering the complete transit event along with $\sim$55 minutes of pre-ingress and $\sim$65 minutes of post-egress. The weather condition was mostly clear, and the seeing varied between 1$\farcs$2 and 1$\farcs$6 with a median value of 1$\farcs$4 at around 7500~\AA, corresponding to a seeing-limited resolution of 3.4--4.5~\AA. We adopted an exposure time of 30 seconds for the blue channel and 180 seconds for the red channel, and collected a total of 344 (blue) and 88 (red) frames. The spectra of FeAr and HeNeAr arc lamps were acquired through a 0$\farcs$5-wide slit.

%++++++++++++++++++++++++++++++++++
%   Figure
%++++++++++++++++++++++++++++++++++
\begin{figure*}
\centering
\includegraphics[width=\textwidth]{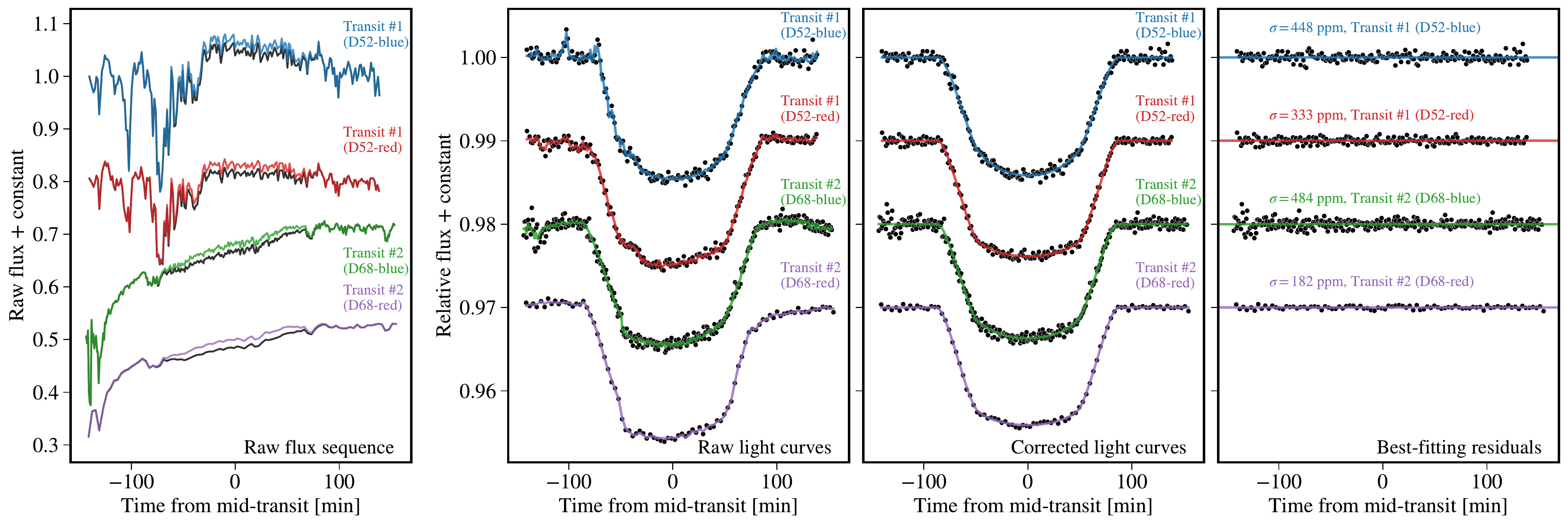}
\caption{White light curves for the two transits observed by P200/DBSP. The {\it left panel} presents the raw flux time series of the target and reference stars. The three {\it right panels} present the reference-calibrated light curves: before systematics correction, after systematics correction, and best-fitting residuals. }
\label{fig:dbsp_wlc}
\end{figure*}

%------------------------------------------------------------------------
%  3. DATA REDUCTION
%------------------------------------------------------------------------
\section{Data reduction}
\label{sec:datared}

We reduced the P200/DBSP raw data based on our established long-slit transit spectroscopy pipeline, which was first developed to reduce the Gran Telescopio Canarias (GTC) long-slit spectroscopic data \citep[e.g.,][]{2017A&A...600L..11C,2017A&A...600A.138C,2018A&A...616A.145C,2020A&A...642A..54C,2021ApJ...913L..16C} and then adapted to reduce the DBSP data in \citet{2021MNRAS.500.5420C}. For completeness, we give below a brief summary of our data reduction procedures.

The two-dimensional (2D) raw spectral images were calibrated for overscan, bias, flat field, sky, and cosmic rays. The one-dimensional (1D) spectra were extracted using the \texttt{APALL} task of \texttt{IRAF} \citep{1986SPIE..627..733T,1993ASPC...52..173T} with the optimal extraction algorithm \citep{1986PASP...98..609H}. The wavelength solutions were constructed from the arc lines using the fourth-order B-spline function, implemented by the IDL script \texttt{bspline\_iterfit.pro} written by D. Schlegel and S. Burles. The spectral drifts over time were determined by measuring stellar line centers in each spectrum, and the misalignments between target and reference stars were calculated through the cross-correlation of telluric absorption lines. The wavelength solutions were shifted to match the rest-frame air values to correct for the offsets resulting from any imperfect alignment to the slit center. The 1D spectra were kept in the original pixel space without linearization, while the drifts and misalignments were corrected in the wavelength solutions. To create the white and spectroscopic light curves, the mid-exposure time was extracted and converted to Barycentric Julian Dates in Barycentric Dynamical Time \citep[$\mathrm{BJD}_\mathrm{TDB}$;][]{2010PASP..122..935E}, and the flux was then integrated within a certain pixel range which was converted from the requested bandpass using the wavelength solutions. 

In contrast to our previous studies where we selected the best aperture radius among a wide range of sizes, here we restrained ourselves from large aperture radii due to the small separation of 11.3$''$ between the target and reference stars. For both observing runs, we adopted an aperture radius of 4 pixels in the blue channel and 5.5 pixels in the red channel, equivalent to roughly 1.6$''$ for both channels. This choice was a compromise between light-curve scatter minimization and dilution minimization. Larger aperture radii would increase the reference star's flux inside the target star's aperture, and vice versa, which would result in a diluted transit signal in the target light curve and a leaked transit signal in the reference light curve. 

To estimate the dilution, we adopted an approach similar to that of \citet{2019AA...631A.169T}, in which we mirrored the undiluted part of the diluting star's PSF to remove the diluting star and performed a simple spectral extraction using the same spectral trace of the original star. Due to seeing variation, the dilution is expected to show temporal variation. When the planet was fully inside the stellar disk, we measured a dilution to target flux of $<$0.04\% (D52-blue), $<$0.16\% (D52-red), $<$0.08\% (D68-blue), $<$0.03\% (D68-red) for the white light curves. The dilution was relatively flat in most wavelengths except for $\lambda>9500$~\AA, where a steep rise occurred and the maximum dilution appeared at the longest wavelengths ($<$0.5\% for D52-red and $<$0.2\% for D68-red) in the exposure with the maximum seeing. We experimented with and without dilution correction in the light curve creation to inspect its impact on the transmission spectrum. We found that such correction would cause a typical change in transit depth at 0.14$\sigma$, much smaller than the 1$\sigma$ uncertainty but with noisier behavior. Therefore, we did not perform the dilution correction on our data.

Figure \ref{fig:dbsp_spec} gives an example of the extracted 1D stellar spectra for the target and reference stars. To create the white light curves, the target and reference fluxes were individually integrated in the wavelength ranges of 3250--5018~\AA~(D52-blue), 6168--10007~\AA~(D52-red), 4168--6768~\AA~(D68-blue), and 6778--10007~\AA~~(D68-red), respectively. The target flux time series were then divided by that of the reference and then normalized by the out-of-transit level. To create the spectroscopic light curves, similar procedures were performed in the designed narrow bands. The full wavelength range was divided into 38 narrow bands, most of which had a width of 200~{\AA} while two groups of five bands around the Na and K wavelengths had a width of 30~{\AA} and 50~\AA, respectively. The four white light curves are shown in Figure~\ref{fig:dbsp_wlc}, and the spectroscopic light curves are shown in Appendix \ref{sec:app_fig}.

%++++++++++++++++++++++++++++++++++
%   Table
%++++++++++++++++++++++++++++++++++
\begin{deluxetable*}{lcrrrr}
\tablecaption{Priors and posteriors of the parameter estimation for white light curves. \label{tab:lcparams}}
\tablewidth{0pt}
\tablecolumns{6}
\tablehead{
\colhead{Parameter} & 
\colhead{Prior} & 
\multicolumn{4}{c}{Posterior estimate}
}
\startdata
    $i$ [$^{\circ}$]                    & $\mathcal{U}(80,90)$           & \multicolumn{4}{c}{$85.652^{+0.058}_{-0.058}$} \\ \noalign{\smallskip}
    $a/R_\star$                         & $\mathcal{U}(7,12)$            & \multicolumn{4}{c}{$9.864^{+0.069}_{-0.069}$} \\ \noalign{\smallskip}
    \hline\noalign{\smallskip}
    & & \multicolumn{2}{c}{Transit 1} &  \multicolumn{2}{c}{Transit 2} \\ \noalign{\smallskip}
    $T_\mathrm{mid}$ [$\mathrm{BJD}_\mathrm{TDB}-2450000$] & $\mathcal{U}(6185.770,6185.810)$  & \multicolumn{2}{c}{$6185.79050^{+0.00015}_{-0.00015}$} & 
                                                                                  \multicolumn{2}{c}{--} 
                                                                                  \\ \noalign{\smallskip}
    $T_\mathrm{mid}$ [$\mathrm{BJD}_\mathrm{TDB}-2450000$] & $\mathcal{U}(7596.805,7596.845)$  & \multicolumn{2}{c}{--} & 
                                                                                  \multicolumn{2}{c}{$7596.82464^{+0.00016}_{-0.00016}$} 
                                                                                  \\ \noalign{\smallskip}
    \hline\noalign{\smallskip}
    & & D52-blue & D52-red & D68-blue & D68-red \\ \noalign{\smallskip}
    $R_\mathrm{p}/R_\star$              & $\mathcal{U}(0.07,0.17)$       &  $0.1189^{+0.0008}_{-0.0008}$ 
                                                                         &  $0.1171^{+0.0015}_{-0.0015}$
                                                                         &  $0.1164^{+0.0014}_{-0.0014}$
                                                                         &  $0.1182^{+0.0013}_{-0.0013}$\\ \noalign{\smallskip}
    $\sigma_\mathrm{j}$ [$10^{-6}$]     & $\mathcal{U}(0.1,5000)$        &  $417^{+28}_{-28}$ 
                                                                         &  $341^{+20}_{-20}$
                                                                         &  $459^{+25}_{-25}$
                                                                         &  $186^{+21}_{-21}$\\ \noalign{\smallskip}
    $\ln A$                             & $\mathcal{U}(-10,-1)$          &  $-6.2^{+0.6}_{-0.6}$ 
                                                                         &  $-7.2^{+0.7}_{-0.6}$
                                                                         &  $-7.8^{+0.4}_{-0.3}$
                                                                         &  $-6.7^{+0.8}_{-0.6}$\\ \noalign{\smallskip}
    $\ln\tau_t$                         & $\mathcal{U}(-6,5)$            &  $1.4^{+2.3}_{-1.5}$ 
                                                                         &  $-1.8^{+0.6}_{-0.5}$
                                                                         &  $0.9^{+2.7}_{-2.5}$
                                                                         &  $-0.7^{+0.8}_{-0.8}$\\ \noalign{\smallskip}
    $\ln\tau_x$                         & $\mathcal{U}(-5,5)$            &  $4.0^{+0.6}_{-0.9}$ 
                                                                         &  $3.9^{+0.8}_{-1.2}$
                                                                         &  $0.2^{+0.7}_{-0.6}$
                                                                         &  $2.1^{+0.8}_{-0.7}$\\ \noalign{\smallskip}
    $\ln\tau_y$                         & $\mathcal{U}(-5,5)$            &  $3.8^{+0.8}_{-1.0}$ 
                                                                         &  $2.9^{+1.3}_{-1.1}$
                                                                         &  $1.0^{+0.6}_{-0.6}$
                                                                         &  $3.1^{+1.1}_{-0.9}$\\ \noalign{\smallskip}
    $\ln\tau_s$                         & $\mathcal{U}(-5,5)$            &  $2.4^{+0.6}_{-0.6}$ 
                                                                         &  $2.2^{+0.7}_{-0.8}$
                                                                         &  $2.0^{+0.4}_{-0.4}$
                                                                         &  $3.3^{+0.9}_{-0.8}$\\ \noalign{\smallskip}
    $u_1$                               & $\mathcal{N}(0.607,0.063^2)$   &  $0.59^{+0.04}_{-0.04}$ & -- & -- & --\\ \noalign{\smallskip}
    $u_2$                               & $\mathcal{N}(0.198,0.048^2)$   &  $0.20^{+0.04}_{-0.04}$ & -- & -- & --\\ \noalign{\smallskip}
    $u_1$                               & $\mathcal{N}(0.281,0.036^2)$   & -- & $0.27^{+0.03}_{-0.03}$ & -- & --\\ \noalign{\smallskip}
    $u_2$                               & $\mathcal{N}(0.294,0.018^2)$   & -- & $0.29^{+0.02}_{-0.02}$ & -- & --\\ \noalign{\smallskip}
    $u_1$                               & $\mathcal{N}(0.410,0.047^2)$   & -- & -- & $0.38^{+0.03}_{-0.03}$ & --\\ \noalign{\smallskip}
    $u_2$                               & $\mathcal{N}(0.286,0.028^2)$   & -- & -- & $0.28^{+0.03}_{-0.03}$ & --\\ \noalign{\smallskip}
    $u_1$                               & $\mathcal{N}(0.256,0.033^2)$   & -- & -- & -- & $0.25^{+0.03}_{-0.03}$ \\ \noalign{\smallskip}
    $u_2$                               & $\mathcal{N}(0.290,0.016^2)$   & -- & -- & -- & $0.29^{+0.02}_{-0.02}$ \\ \noalign{\smallskip}
\enddata
\end{deluxetable*}

%------------------------------------------------------------------------
%  4. LIGHT-CURVE ANALYSIS
%------------------------------------------------------------------------
\section{Light-curve analysis}
\label{sec:lc}

As in our previous work \citep{2018A&A...616A.145C,2020A&A...642A..54C,2021MNRAS.500.5420C,2021ApJ...913L..16C}, we treated the observed light curves as Gaussian processes \citep[GP;][]{2006gpml.book.....R}, which was first introduced to model transmission spectroscopy by \citet{2012MNRAS.419.2683G}. Both white and spectroscopic light curves are expected to be composed of the transit signal and correlated systematics. The python package \texttt{george} \citep{2015ITPAM..38..252A} was employed to implement the GP, where the transit model was adopted as the mean function and the Mat\'{e}rn $\nu=3/2$ kernel function was adopted as the GP covariance matrix. 

The transit model was implemented by \texttt{batman} \citep{2015PASP..127.1161K} with the quadratic limb-darkening law, which was parameterized by orbital inclination ($i$), scaled semi-major axis ($a/R_\star$), planet-to-star radius ratio ($R_\mathrm{p}/R_\star$), mid-transit time ($T_\mathrm{mid}$), and limb-darkening coefficients ($u_1$, $u_2$). The orbital period ($P$) was fixed to 4.46529976~d from \citet{2014MNRAS.437...46N}, while the eccentricity was fixed to zero. 

For the GP covariance matrix, multiple state vectors were considered as the input variables, including time ($t$), spectral drifts ($x$), spatial drifts ($y$), and seeing ($s$; as represented by spatial FWHM). The latter three have been standardized, i.e., subtracted by the mean value and then divided by the standard deviation. The associated free parameters were the covariance amplitude ($A$) and multiple correlation length scales ($\tau_k$, where $k=\{t,x,y,s\}$). To account for potential underestimation of white noise, a free parameter $\sigma_\mathrm{j}$ was added in quadrature to the propagated photon-noise dominated flux uncertainties. 

During parameter estimation, uniform or log-uniform priors were adopted for all the parameters, except for the limb-darkening coefficients. Normal priors were imposed on the limb-darkening coefficients, with the mean and width values being prepared in advance, which were calculated from the ATLAS9 stellar atmosphere models using a python script written by \citet{2015MNRAS.450.1879E}. The stellar model grid with $T_\star=6000$~K, $\log g=4.5$, and $\mathrm{[Fe/H]}=0.0$ was used to derive the mean values, while the grids with $T_\star=5750$~K and $T_\star=6250$~K were used to calculate a mean difference from that with $T_\star=6000$~K as the width values.

The posterior distributions of free parameters were explored by the affine-invariant Markov chain Monte Carlo (MCMC) as implemented by \texttt{emcee} \citep{2013PASP..125..306F}. When model comparison was necessary, the Bayesian multimodal nested sampling algorithm \citep{2009MNRAS.398.1601F} was applied using \texttt{PyMultiNest} \citep{2014A&A...564A.125B}.

%------------------------------------------------------------------------
\subsection{White light curves}
\label{sec:wlc}

We jointly fitted the four white light curves, which shared the same values of $i$ and $a/R_\star$ for all and the same values of $T_\mathrm{mid}$ for the same transit. For both transits, we adopted all four inputs \{$t$, $x$, $y$, $s$\} in the GP covariance matrix, which showed the highest Bayesian evidence as calculated by \texttt{PyMultiNest} among the experiments with either single input or different combinations of them. We used \texttt{emcee} to explore the posterior distributions of the 40 free parameters. Three runs of 120 walkers were performed, with the first two in 1000 steps for initial burn-in and the third in 50000 steps (but thinned every ten) for final production.

Figure \ref{fig:dbsp_wlc} shows the best-fit light-curve models compared to the observations. The root mean square (rms) of the best-fit light-curve residuals reach 448, 333, 484, 182 parts per million (ppm), for D52-blue, D52-red, D68-blue, D68-red, respectively, which correspond to 2.32$\times$, 4.72$\times$, 2.42$\times$, 2.28$\times$ photon noise. Table \ref{tab:lcparams} gives the priors and posteriors for the free parameters in the joint analysis of white light curves. Our derived values of $i$ and $a/R_\star$ agree well with those of \citet{2014MNRAS.437...46N}, i.e., $i=85\fdg634\pm0\fdg056$ and $a/R_\star=9.853\pm0.071$.

%++++++++++++++++++++++++++++++++++
%   Table
%++++++++++++++++++++++++++++++++++
\begin{deluxetable}{rcrc}
\tablecaption{Mid-transit times of HAT-P-1b. \label{tab:tmid}}
\tablewidth{0pt}
\tablecolumns{4}
\tablehead{
\colhead{Epoch} & 
\colhead{Mid-transit} & 
\colhead{O$-$C} &
\colhead{Reference}
\\
\colhead{} & 
\colhead{($\mathrm{BJD}_\mathrm{TDB}-2450000$)} & 
\colhead{(sec)} &
\colhead{}
}
\startdata
$-$517 & 3979.930708 $\pm$ 0.000690 & $-$121.9 & 2\\
$-$516 & 3984.397768 $\pm$ 0.009000 & 30.2 & 1\\
$-$515 & 3988.862738 $\pm$ 0.000760 & 1.8 & 2\\
$-$513 & 3997.792768 $\pm$ 0.000540 & $-$47.4 & 2\\
$-$513 & 3997.794248 $\pm$ 0.000470 & 80.4 & 2\\
$-$511 & 4006.724028 $\pm$ 0.000590 & 9.7 & 2\\
$-$509 & 4015.654147 $\pm$ 0.001070 & $-$31.8 & 2\\
$-$497 & 4069.238716 $\pm$ 0.002900 & 52.3 & 2\\
$-$431 & 4363.946769 $\pm$ 0.000910 & $-$96.5 & 3\\
$-$427 & 4381.809249 $\pm$ 0.001250 & 14.2 & 3\\
$-$48 & 6074.157370 $\pm$ 0.000180 & $-$23.4 & 4\\
$-$47 & 6078.623070 $\pm$ 0.000180 & 11.2 & 4\\
$-$39 & 6114.345370 $\pm$ 0.000200 & 2.8 & 4\\
$-$30 & 6154.532530 $\pm$ 0.000180 & $-$43.6 & 5\\
$-$23 & 6185.790501 $\pm$ 0.000153 & 31.9 & 8\\
$-$22 & 6190.255420 $\pm$ 0.000180 & $-$1.0 & 4\\
$-$19 & 6203.649839 $\pm$ 0.000950 & $-$128.8 & 6\\
$-$15 & 6221.512610 $\pm$ 0.000220 & 7.0 & 5\\
$-$10 & 6243.839084 $\pm$ 0.000076 & 5.0 & 7\\
233\tablenotemark{\footnotesize a} & 7328.907659 $\pm$ 0.000034 & 71.2 & 7\\
293 & 7596.824638 $\pm$ 0.000163 & $-$15.0 & 8\\
549 & 8739.942711 $\pm$ 0.000667 & 103.3 & 9\\
550 & 8744.407173 $\pm$ 0.001247 & 31.0 & 9\\
551 & 8748.873683 $\pm$ 0.000769 & 135.6 & 9\\
552 & 8753.337067 $\pm$ 0.000697 & $-$30.0 & 9\\
553 & 8757.802413 $\pm$ 0.000708 & $-$25.9 & 9\\
554 & 8762.267168 $\pm$ 0.000593 & $-$73.0 & 9
\enddata
\tablenotetext{a}{This epoch was not used in the linear regression. It was listed here for completeness.}
\tablecomments{The zero epoch and orbital period are revised as $T_0(\mathrm{BJD}_\mathrm{TDB})=2456288.492023\pm0.000063$ and $P=4.46529962\pm0.00000039$~d, respectively.}
\tablerefs{1. \citet{2007ApJ...656..552B}; 2. \citet{2007AJ....134.1707W}; 3. \citet{2008ApJ...686..649J}; 4. \citet{2014MNRAS.437...46N}; 5. \citet{2015ApJ...811...55M}; 6. \citet{2016MNRAS.459..789T}; 7. \citet{2019AA...631A.169T}; 8. This work (DBSP); 9. This work (TESS).}
\end{deluxetable}

%++++++++++++++++++++++++++++++++++
%   Figure
%++++++++++++++++++++++++++++++++++
\begin{figure}
\centering
\includegraphics[width=\columnwidth]{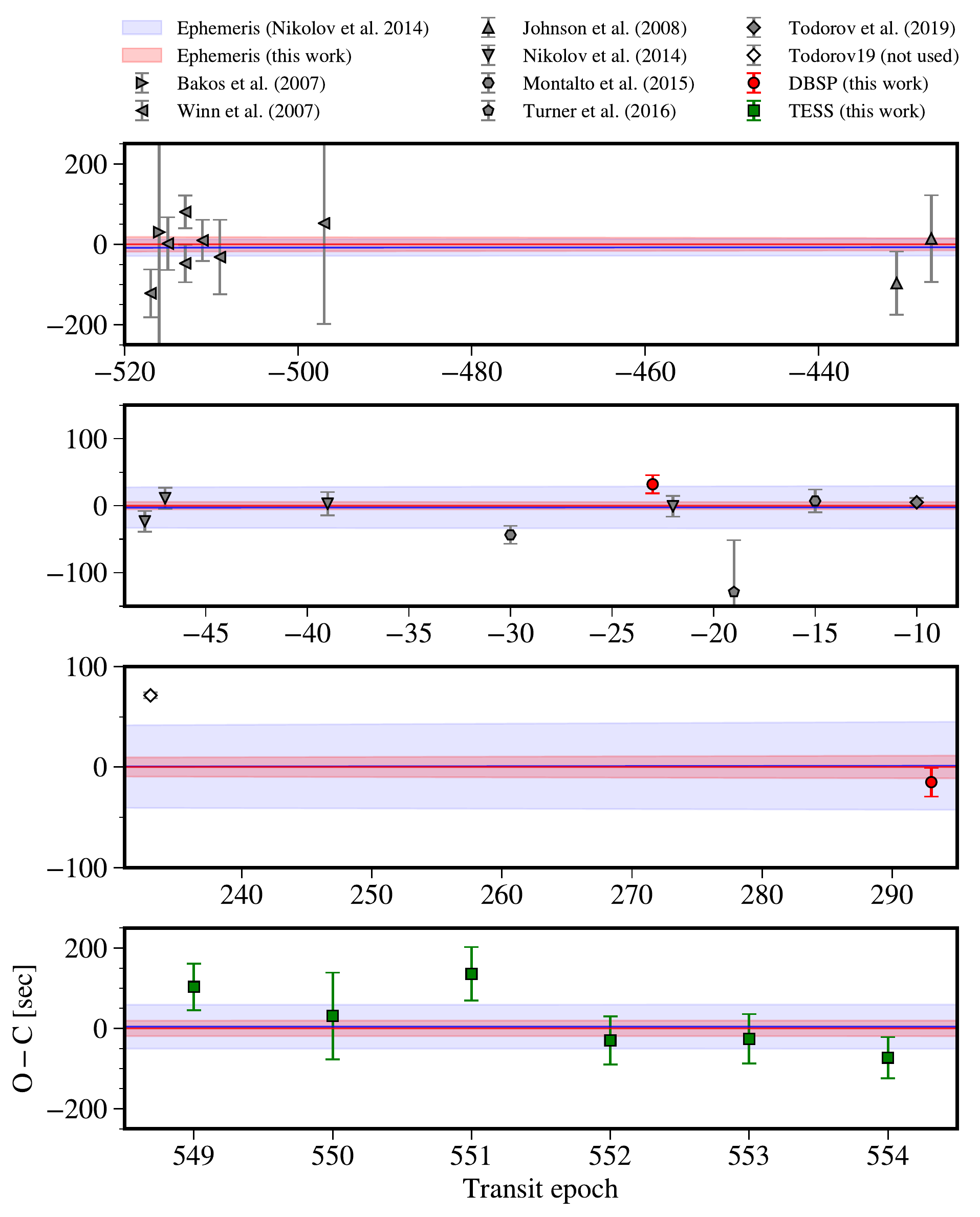}
\caption{Residuals of the mid-transit times after a linear ephemeris regression. For clarity, the epochs are grouped into four rows. The red line and shaded region present the best-fit model and 1$\sigma$ confidence region from this work, while the blue shows that derived from \citet{2014MNRAS.437...46N}.}
\label{fig:eph}
\end{figure}

%------------------------------------------------------------------------
\subsubsection{Update of transit ephemeris}
\label{sec:eph}

With our two newly measured transit timings, we attempted to revise the transit ephemeris of HAT-P-1b. We collected other transit timing measurements from  \citet{2007ApJ...656..552B}, \citet{2007AJ....134.1707W}, \citet{2008ApJ...686..649J}, \citet{2014MNRAS.437...46N}, \citet{2015ApJ...811...55M},  \citet{2016MNRAS.459..789T}, and \citet{2019AA...631A.169T}, and converted them into $\mathrm{BJD}_\mathrm{TDB}$ using the online applet\footnote{\url{https://astroutils.astronomy.osu.edu/time/}} developed by \citet{2010PASP..122..935E} if a different time standard was adopted. We did not include the transit timings of \citet{2015MNRAS.450..192W} since the reported best-fit values did not have sufficient significant digits. We also excluded the transit timing measured by Gemini GMOS B600 presented in \citet{2019AA...631A.169T}, which was derived from a light curve with strong systematics and the uncertainty was likely underestimated.

We noticed that HAT-P-1 was observed by the Transiting Exoplanet Survey Satellite \citep[TESS;][]{2014SPIE.9143E..20R} in Sector 16 between 2019 September 12 and October 6. The data were collected in full frame images (FFIs) with a cadence of 30 minutes. In order to expand the time span and update the transit ephemeris, we downloaded the light curve produced by the MIT Quick Look Pipeline \citep{2020RNAAS...4..204H,2020RNAAS...4..206H}. For the subsequent timing measurements, we adopted the detrended flux (labeled as \texttt{KSPSAP\_FLUX}). For each transit, we extracted an individual light curve in a window of three times transit duration on both sides of each expected mid-transit time. 

The TESS light curves were modeled in an approach similar to DBSP. In the transit model, we fixed the values of $i$ and $a/R_\star$ to those listed in Table \ref{tab:lcparams}, and adopted the nonlinear limb-darkening law with the coefficients fixed to pre-calculated values ($c_1=0.5404$, $c_2=0.0740$, $c_3=0.2020$, $c_4=-0.1518$). The radius ratio was fitted with a normal prior $\mathcal{N}(0.11802,0.00018^2)$. A dilution parameter was freely fitted to correct for potential imperfect dilution correction in the MIT Quick Look Pipeline. For the GP covariance matrix, only time was adopted as the single input. The long cadence has been considered by first generating a super-sampled model and then binning it into current cadence. 

Table \ref{tab:tmid} gives the full list of literature transit timings, two DBSP timings, and six TESS timings. We employed \texttt{emcee} to perform a linear regression to the transit epochs ($E$) and timings ($T_\mathrm{C}$) to derive the zero epoch timing ($T_0$) and orbital period ($P$): $T_\mathrm{C}=T_0+EP$. We used a free parameter to rescale the timing uncertainties to account for potential underestimation. The zero epoch was chosen as the one that minimized the uncertainty of $T_0$. Finally, we derived 
\begin{eqnarray}
T_0 &=& 2456288.492023 \pm 0.000063~\mathrm{BJD}_\mathrm{TDB}, \\
P &=& 4.46529962 \pm 0.00000039~\mathrm{d},
\end{eqnarray}
which are consistent with that of \citet{2014MNRAS.437...46N} but with smaller uncertainties. The derived uncertainty rescaling factor is $1.3\pm0.2$. Figure \ref{fig:eph} shows the comparison between our newly derived transit ephemeris and that of \citet{2014MNRAS.437...46N}.

%++++++++++++++++++++++++++++++++++
%   Figure
%++++++++++++++++++++++++++++++++++
\begin{figure*}
\centering
\includegraphics[width=\textwidth]{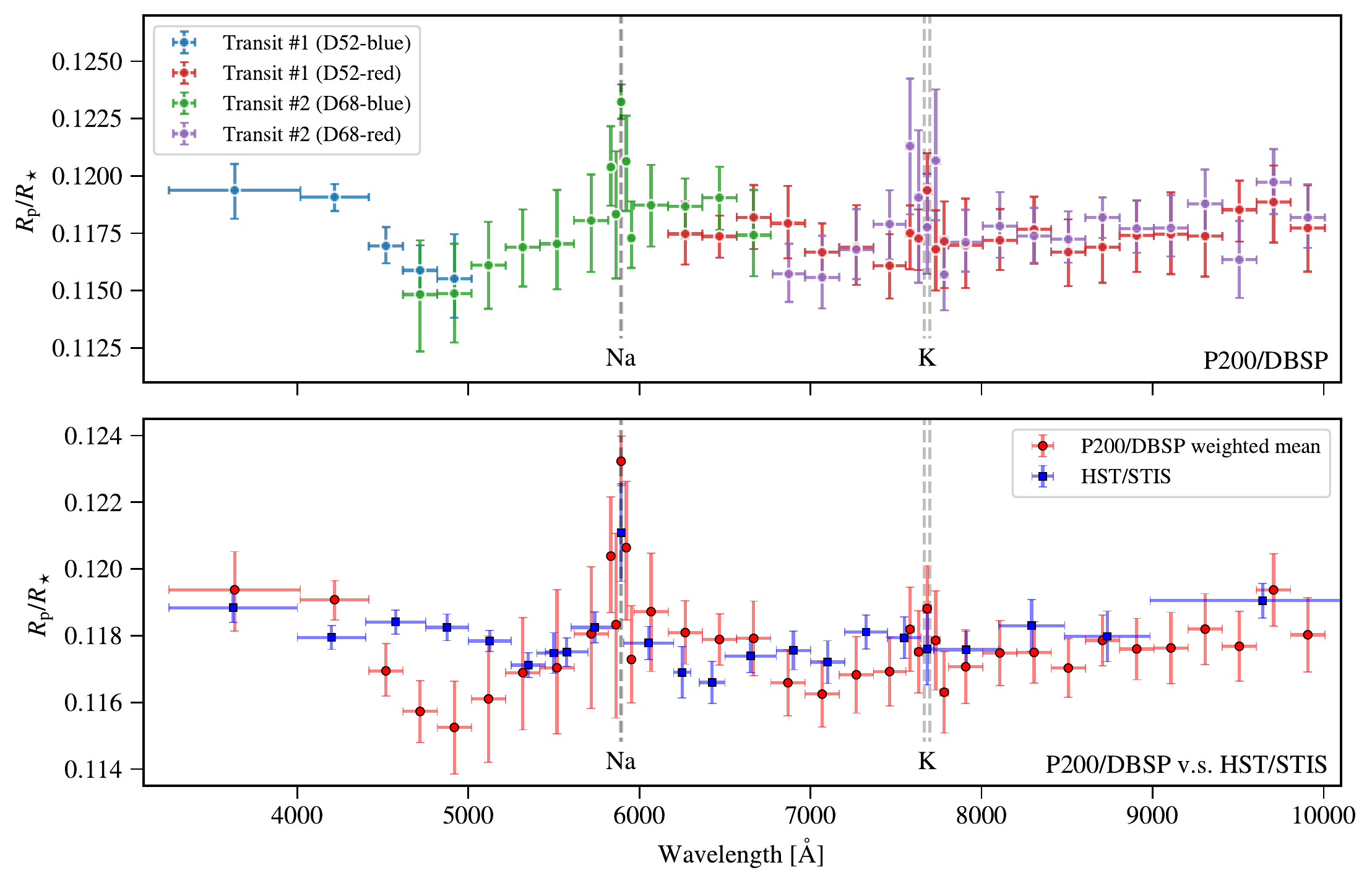}
\caption{{\it Top panel:} comparison between the transmission spectra acquired by P200/DBSP in two transits. {\it Bottom panel:} comparison between the combined P200/DBSP transmission spectrum and the HST/STIS transmission spectrum. The vertical dashed lines mark the locations of the Na and K doublets.}
\label{fig:dbsp_ts_2nights}
\end{figure*}

%++++++++++++++++++++++++++++++++++
%   Figure
%++++++++++++++++++++++++++++++++++
\begin{figure}
\centering
\includegraphics[width=\columnwidth]{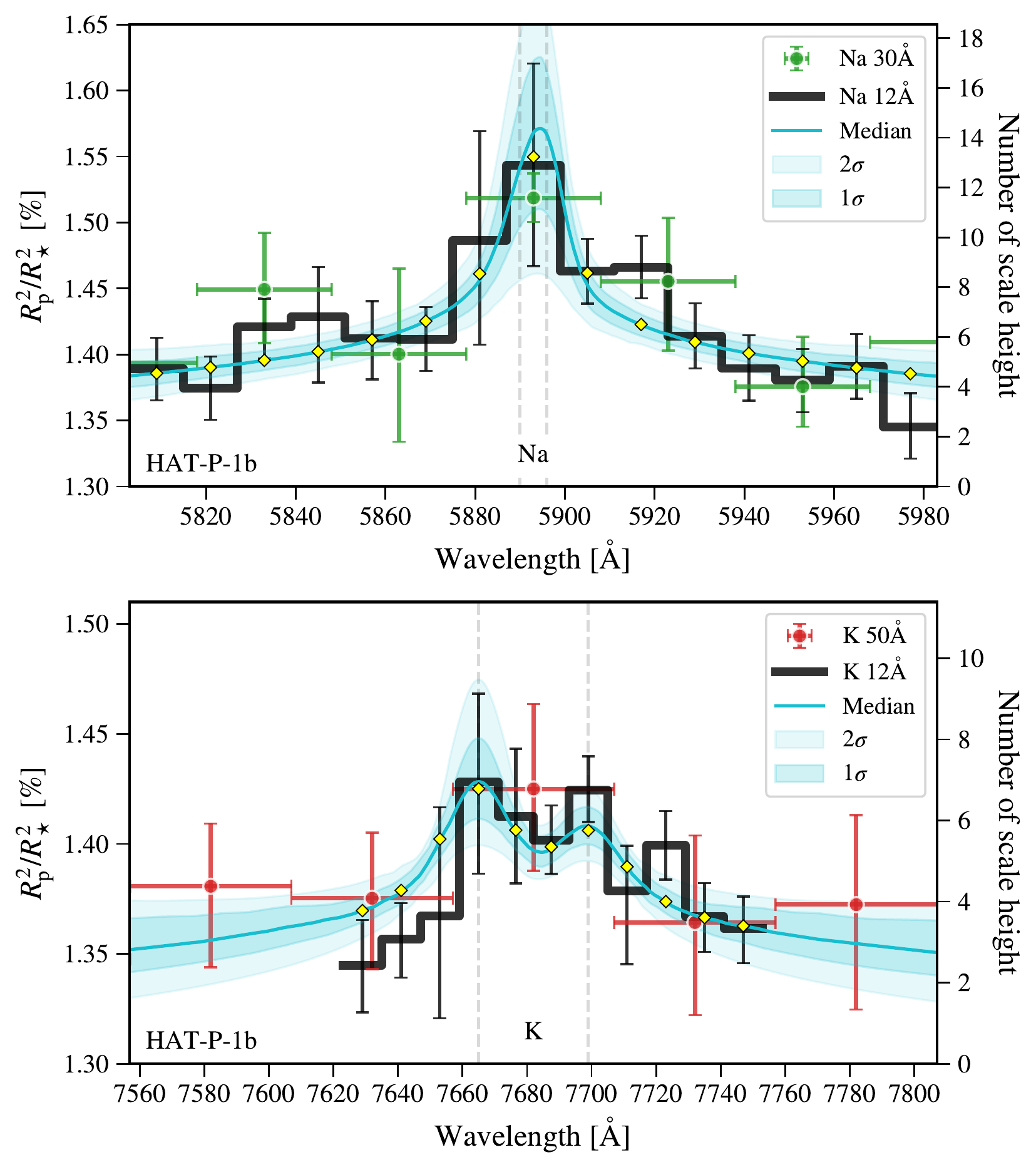}
\caption{Ultra-narrow band transmission spectra around the Na (top) and K (bottom) lines. The cyan line and shaded regions show the median model and 1$\sigma$/2$\sigma$ confidence regions from the Bayesian spectral retrieval. Also shown for comparison are the 30~\AA\ (Na) and 50~\AA\ (K) bands in the full P200/DBSP spectrum.}
\label{fig:dbsp_NaK}
\end{figure}

%++++++++++++++++++++++++++++++++++
%   Table
%++++++++++++++++++++++++++++++++++
\begin{deluxetable*}{ccccccccc}
\tablecaption{Derived transmission spectrum of HAT-P-1b.\label{tab:ts}}
\tablewidth{0pt}
\tablecolumns{9}
\tabletypesize{\footnotesize}
\tablehead{
\colhead{$\lambda$ (\AA)} & 
\colhead{$u_1$ prior} & 
\colhead{$u_2$ prior} &
\colhead{$R_\mathrm{p}/R_\star$} &
\multicolumn{2}{c}{$R_\mathrm{p}/R_\star$} &
\colhead{} &
\multicolumn{2}{c}{RMS/Photon noise}
\\
\cline{5-6}\cline{8-9}
\colhead{} &
\colhead{} &
\colhead{} &
\colhead{Weighted} &
\colhead{Transit 1} &
\colhead{Transit 2} &
\colhead{} &
\colhead{Transit 1} &
\colhead{Transit 2}
} 
\startdata
    3250--4018 & $\mathcal{N}(0.735,0.095^2)$ & $\mathcal{N}(0.108,0.083^2)$ & $0.1194 ^{+0.0011}_{-0.0012}$ & $0.1194 ^{+0.0011}_{-0.0012}$ & -- & & 1.41 & --\\
    4018--4418 & $\mathcal{N}(0.677,0.071^2)$ & $\mathcal{N}(0.148,0.059^2)$ & $0.1191 ^{+0.0006}_{-0.0006}$ & $0.1191 ^{+0.0006}_{-0.0006}$ & -- & & 1.38 & --\\
    4418--4618 & $\mathcal{N}(0.584,0.059^2)$ & $\mathcal{N}(0.220,0.042^2)$ & $0.1169 ^{+0.0008}_{-0.0008}$ & $0.1169 ^{+0.0008}_{-0.0008}$ & -- & & 1.37 & --\\
    4618--4818 & $\mathcal{N}(0.538,0.058^2)$ & $\mathcal{N}(0.247,0.039^2)$ & $0.1157 ^{+0.0009}_{-0.0009}$ & $0.1159 ^{+0.0010}_{-0.0010}$ & $0.1148 ^{+0.0024}_{-0.0024}$ & & 1.52 & 1.17\\
    4818--5018 & $\mathcal{N}(0.482,0.060^2)$ & $\mathcal{N}(0.275,0.040^2)$ & $0.1152 ^{+0.0014}_{-0.0014}$ & $0.1155 ^{+0.0018}_{-0.0018}$ & $0.1149 ^{+0.0021}_{-0.0021}$ & & 1.19 & 1.16\\
    5018--5218 & $\mathcal{N}(0.483,0.057^2)$ & $\mathcal{N}(0.259,0.038^2)$ & $0.1161 ^{+0.0019}_{-0.0019}$ & -- & $0.1161 ^{+0.0019}_{-0.0019}$ & & -- & 1.20\\
    5218--5418 & $\mathcal{N}(0.451,0.050^2)$ & $\mathcal{N}(0.273,0.031^2)$ & $0.1169 ^{+0.0017}_{-0.0017}$ & -- & $0.1169 ^{+0.0017}_{-0.0017}$ & & -- & 1.26\\
    5418--5618 & $\mathcal{N}(0.426,0.049^2)$ & $\mathcal{N}(0.283,0.029^2)$ & $0.1170 ^{+0.0023}_{-0.0020}$ & -- & $0.1170 ^{+0.0023}_{-0.0020}$ & & -- & 1.32\\
    5618--5818 & $\mathcal{N}(0.405,0.046^2)$ & $\mathcal{N}(0.290,0.026^2)$ & $0.1181 ^{+0.0020}_{-0.0022}$ & -- & $0.1181 ^{+0.0020}_{-0.0022}$ & & -- & 1.22\\
    5818--5848 & $\mathcal{N}(0.389,0.046^2)$ & $\mathcal{N}(0.299,0.026^2)$ & $0.1204 ^{+0.0018}_{-0.0017}$ & -- & $0.1204 ^{+0.0018}_{-0.0017}$ & & -- & 1.09\\
    5848--5878 & $\mathcal{N}(0.387,0.045^2)$ & $\mathcal{N}(0.297,0.025^2)$ & $0.1183 ^{+0.0027}_{-0.0028}$ & -- & $0.1183 ^{+0.0027}_{-0.0028}$ & & -- & 1.04\\
    5878--5908 & $\mathcal{N}(0.392,0.046^2)$ & $\mathcal{N}(0.286,0.028^2)$ & $0.1232 ^{+0.0008}_{-0.0007}$ & -- & $0.1232 ^{+0.0008}_{-0.0007}$ & & -- & 1.11\\
    5908--5938 & $\mathcal{N}(0.382,0.045^2)$ & $\mathcal{N}(0.298,0.025^2)$ & $0.1206 ^{+0.0020}_{-0.0022}$ & -- & $0.1206 ^{+0.0020}_{-0.0022}$ & & -- & 1.03\\
    5938--5968 & $\mathcal{N}(0.379,0.045^2)$ & $\mathcal{N}(0.299,0.025^2)$ & $0.1173 ^{+0.0016}_{-0.0013}$ & -- & $0.1173 ^{+0.0016}_{-0.0013}$ & & -- & 1.04\\
    5968--6168 & $\mathcal{N}(0.370,0.043^2)$ & $\mathcal{N}(0.297,0.023^2)$ & $0.1187 ^{+0.0018}_{-0.0018}$ & -- & $0.1187 ^{+0.0018}_{-0.0018}$ & & -- & 1.33\\
    6168--6368 & $\mathcal{N}(0.355,0.042^2)$ & $\mathcal{N}(0.297,0.022^2)$ & $0.1181 ^{+0.0009}_{-0.0009}$ & $0.1175 ^{+0.0014}_{-0.0014}$ & $0.1187 ^{+0.0013}_{-0.0013}$ & & 1.58 & 1.19\\
    6368--6568 & $\mathcal{N}(0.322,0.043^2)$ & $\mathcal{N}(0.309,0.021^2)$ & $0.1179 ^{+0.0008}_{-0.0008}$ & $0.1174 ^{+0.0009}_{-0.0009}$ & $0.1191 ^{+0.0014}_{-0.0014}$ & & 1.59 & 1.23\\
    6568--6768 & $\mathcal{N}(0.311,0.041^2)$ & $\mathcal{N}(0.308,0.021^2)$ & $0.1179 ^{+0.0011}_{-0.0011}$ & $0.1182 ^{+0.0014}_{-0.0014}$ & $0.1174 ^{+0.0019}_{-0.0019}$ & & 1.55 & 1.15\\
    6768--6968 & $\mathcal{N}(0.317,0.039^2)$ & $\mathcal{N}(0.296,0.020^2)$ & $0.1166 ^{+0.0010}_{-0.0010}$ & $0.1179 ^{+0.0016}_{-0.0016}$ & $0.1157 ^{+0.0013}_{-0.0013}$ & & 1.58 & 1.15\\
    6968--7168 & $\mathcal{N}(0.305,0.038^2)$ & $\mathcal{N}(0.295,0.019^2)$ & $0.1162 ^{+0.0010}_{-0.0010}$ & $0.1167 ^{+0.0013}_{-0.0013}$ & $0.1156 ^{+0.0016}_{-0.0016}$ & & 1.39 & 1.10\\
    7168--7368 & $\mathcal{N}(0.295,0.036^2)$ & $\mathcal{N}(0.293,0.018^2)$ & $0.1168 ^{+0.0011}_{-0.0011}$ & $0.1169 ^{+0.0017}_{-0.0017}$ & $0.1168 ^{+0.0015}_{-0.0015}$ & & 1.54 & 1.31\\
    7368--7557 & $\mathcal{N}(0.283,0.035^2)$ & $\mathcal{N}(0.292,0.017^2)$ & $0.1169 ^{+0.0010}_{-0.0010}$ & $0.1161 ^{+0.0014}_{-0.0014}$ & $0.1179 ^{+0.0015}_{-0.0015}$ & & 1.60 & 1.31\\
    7557--7607 & $\mathcal{N}(0.278,0.035^2)$ & $\mathcal{N}(0.293,0.017^2)$ & $0.1182 ^{+0.0013}_{-0.0013}$ & $0.1175 ^{+0.0014}_{-0.0014}$ & $0.1213 ^{+0.0030}_{-0.0030}$ & & 1.32 & 1.12\\
    7607--7657 & $\mathcal{N}(0.275,0.034^2)$ & $\mathcal{N}(0.293,0.016^2)$ & $0.1175 ^{+0.0012}_{-0.0012}$ & $0.1173 ^{+0.0013}_{-0.0013}$ & $0.1191 ^{+0.0033}_{-0.0033}$ & & 1.21 & 1.46\\
    7657--7707 & $\mathcal{N}(0.272,0.034^2)$ & $\mathcal{N}(0.292,0.016^2)$ & $0.1188 ^{+0.0013}_{-0.0013}$ & $0.1194 ^{+0.0016}_{-0.0016}$ & $0.1178 ^{+0.0022}_{-0.0022}$ & & 1.13 & 1.16\\
    7707--7757 & $\mathcal{N}(0.271,0.034^2)$ & $\mathcal{N}(0.291,0.016^2)$ & $0.1179 ^{+0.0015}_{-0.0015}$ & $0.1168 ^{+0.0017}_{-0.0017}$ & $0.1207 ^{+0.0029}_{-0.0029}$ & & 1.21 & 1.06\\
    7757--7807 & $\mathcal{N}(0.266,0.035^2)$ & $\mathcal{N}(0.294,0.016^2)$ & $0.1163 ^{+0.0012}_{-0.0012}$ & $0.1171 ^{+0.0019}_{-0.0019}$ & $0.1157 ^{+0.0016}_{-0.0016}$ & & 1.12 & 1.07\\
    7807--8007 & $\mathcal{N}(0.263,0.035^2)$ & $\mathcal{N}(0.292,0.016^2)$ & $0.1171 ^{+0.0011}_{-0.0011}$ & $0.1170 ^{+0.0020}_{-0.0020}$ & $0.1171 ^{+0.0013}_{-0.0013}$ & & 1.59 & 1.23\\
    8007--8207 & $\mathcal{N}(0.256,0.035^2)$ & $\mathcal{N}(0.291,0.016^2)$ & $0.1175 ^{+0.0010}_{-0.0010}$ & $0.1172 ^{+0.0013}_{-0.0013}$ & $0.1178 ^{+0.0014}_{-0.0014}$ & & 1.53 & 1.21\\
    8207--8407 & $\mathcal{N}(0.248,0.034^2)$ & $\mathcal{N}(0.289,0.016^2)$ & $0.1175 ^{+0.0009}_{-0.0009}$ & $0.1177 ^{+0.0015}_{-0.0015}$ & $0.1174 ^{+0.0012}_{-0.0012}$ & & 1.55 & 1.26\\
    8407--8607 & $\mathcal{N}(0.233,0.032^2)$ & $\mathcal{N}(0.286,0.014^2)$ & $0.1170 ^{+0.0009}_{-0.0009}$ & $0.1167 ^{+0.0015}_{-0.0015}$ & $0.1172 ^{+0.0011}_{-0.0011}$ & & 1.56 & 1.15\\
    8607--8807 & $\mathcal{N}(0.226,0.032^2)$ & $\mathcal{N}(0.287,0.015^2)$ & $0.1179 ^{+0.0008}_{-0.0008}$ & $0.1169 ^{+0.0015}_{-0.0015}$ & $0.1182 ^{+0.0009}_{-0.0009}$ & & 1.55 & 1.31\\
    8807--9007 & $\mathcal{N}(0.224,0.031^2)$ & $\mathcal{N}(0.287,0.014^2)$ & $0.1176 ^{+0.0009}_{-0.0009}$ & $0.1174 ^{+0.0015}_{-0.0015}$ & $0.1177 ^{+0.0011}_{-0.0011}$ & & 1.57 & 1.12\\
    9007--9207 & $\mathcal{N}(0.221,0.030^2)$ & $\mathcal{N}(0.284,0.013^2)$ & $0.1176 ^{+0.0011}_{-0.0011}$ & $0.1175 ^{+0.0018}_{-0.0018}$ & $0.1177 ^{+0.0013}_{-0.0013}$ & & 1.44 & 1.41\\
    9207--9407 & $\mathcal{N}(0.208,0.030^2)$ & $\mathcal{N}(0.288,0.013^2)$ & $0.1182 ^{+0.0011}_{-0.0011}$ & $0.1174 ^{+0.0016}_{-0.0016}$ & $0.1188 ^{+0.0014}_{-0.0014}$ & & 1.37 & 1.22\\
    9407--9607 & $\mathcal{N}(0.205,0.030^2)$ & $\mathcal{N}(0.287,0.013^2)$ & $0.1177 ^{+0.0010}_{-0.0010}$ & $0.1185 ^{+0.0013}_{-0.0013}$ & $0.1163 ^{+0.0017}_{-0.0017}$ & & 1.37 & 1.18\\
    9607--9807 & $\mathcal{N}(0.216,0.025^2)$ & $\mathcal{N}(0.275,0.010^2)$ & $0.1194 ^{+0.0011}_{-0.0011}$ & $0.1189 ^{+0.0017}_{-0.0017}$ & $0.1197 ^{+0.0014}_{-0.0014}$ & & 1.41 & 1.47\\
   9807--10007 & $\mathcal{N}(0.210,0.025^2)$ & $\mathcal{N}(0.277,0.010^2)$ & $0.1180 ^{+0.0011}_{-0.0011}$ & $0.1177 ^{+0.0019}_{-0.0019}$ & $0.1182 ^{+0.0014}_{-0.0014}$ & & 1.50 & 1.26\\
\enddata
\end{deluxetable*}

%------------------------------------------------------------------------
\subsection{Spectroscopic light curves}
\label{sec:slc}

We fitted the raw spectroscopic light curves individually, without applying the divide-white method to remove the common-mode systematics as in our previous work \citep[e.g.,][]{2021ApJ...913L..16C}. This was mainly because these DBSP spectroscopic light curves came from four ``passbands'' with very different response function (i.e., D52-blue, D52-red, D68-blue, D68-red; see Figure~\ref{fig:dbsp_spec}). According to \citet{2022arXiv220505969J}'s benchmark test on transiting white dwarfs, the divide-white method could potentially introduce an overall offset in the transmission spectrum. With four white light curves from different passbands in hand, it would be difficult to handle the ``offset'' problem if the divide-white method was adopted. 

For each spectroscopic light curve, we fixed the values of $i$, $a/R_\star$, and $T_\mathrm{mid}$ to those listed in Table \ref{tab:lcparams}. We adopted the same four inputs \{$t$, $x$, $y$, $s$\} in the GP covariance matrix as in the white light curve analysis. Consequently, there were nine free parameters for each spectroscopic light curve: $R_\mathrm{p}/R_\star$, $u_1$, $u_2$, $\sigma_\mathrm{j}$, $\ln A$, $\ln\tau_t$, $\ln\tau_x$, $\ln\tau_y$, $\ln\tau_s$. We configured \texttt{emcee} to perform three runs of 32 walkers, with the first two runs in 1000 steps for initial burn-in and the third in 5000 steps for final production.

To produce the final transmission spectrum, we combined the four segments of transmission spectra, and recorded in the overlapping wavelengths the weighted mean along with the propagated uncertainty. Table \ref{tab:ts} presents the adopted priors for the two limb-darkening coefficients, the derived transmission spectra, and the ratio between the rms of best-fit light-curve residuals and photon noise. In general, we achieved an rms of 1.19--1.52$\times$, 1.12--1.60$\times$, 1.03--1.33$\times$, 1.06--1.47$\times$ photon noise in the spectroscopic light curve analysis.

Figure \ref{fig:dbsp_ts_2nights} shows the four segments of DBSP transmission spectra and the final DBSP transmission spectrum compared to the HST/STIS transmission spectrum \citep[taken from][]{2016Natur.529...59S}. The DBSP segments are consistent with each other in the common wavelength ranges. When compared to HST/STIS, the DBSP spectrum agrees with the former in most wavelengths except for $4000\leq\lambda\leq 5000$~\AA. If we fit the DBSP spectrum to the HST/STIS spectrum after the latter is linearly interpolated on the DBSP grid, with an offset as the only free parameter, they would appear consistent with each other at 0.3$\sigma$ confidence level ($\chi^2=29.7$ for 37~degrees of freedom, hereafter dof). On the other hand, the DBSP spectrum is consistent with a flat line only at 4.4$\sigma$ ($\chi^2=85.5$ for 37 dof), indicating that the DBSP spectrum is not flat.

%------------------------------------------------------------------------
\subsection{Ultra-narrow bands in search of Na and K}
In addition to the 38 spectroscopic bands, we also created two groups of ultra-narrow bands, with a bin width of $\sim$12~\AA, centered at the Na and K lines, respectively. The 12~\AA\ bin width is slightly larger than the seeing-limited resolutions (see Table \ref{tab:obsum}), and can be directly compared to the results of \citet{2015MNRAS.450..192W}, in which the GTC tunable 12~\AA\ filters were used. Our Na group consisted of 15 uniform bins centered at 5893~\AA, while our K group consisted of 11 bins centered at 7665~\AA\ and 7699~\AA, of which nine bins had a width of 12~\AA\ and two bins in-between K doublet 11~\AA. 

In our observations, Na was only covered by D68-blue in Transit 2, while K was covered by both transits. However, due to low count rate and sparse cadence of D68-red in Transit 2, we decided to only use the D52-red data for the ultra-narrow band analysis. In this case, we applied the divide-white method to remove the common-mode systematics. We defined 5800-6000~\AA\ and 7600-7800~\AA\ as the local ``white'' bands for Na and K, respectively, which were modeled in the same way as in Section~\ref{sec:slc}. The priors on limb-darkening coefficients were $u_1\sim\mathcal{N}(0.385,0.045^2)$ and $u_2\sim\mathcal{N}(0.296,0.026^2)$ for 5800-6000~\AA\ and $u_1\sim\mathcal{N}(0.271,0.034^2)$ and $u_2\sim\mathcal{N}(0.292,0.016^2)$ for 7600-7800~\AA. We derived $R_\mathrm{p}/R_\star=0.1193\pm0.0014$ for 5800-6000~\AA\ and $R_\mathrm{p}/R_\star=0.1176\pm0.0010$ for 7600-7800~\AA, respectively. 

We derived the common-mode systematics by dividing the local ``white'' light curves by their best-fit transit models, and then removed them from the 12~\AA-bin Na and K ultra-narrow band light curves. Finally, we fitted the corrected Na and K light curves in the same way as in Section~\ref{sec:slc} to derive the Na and K transmission spectra. Figure \ref{fig:dbsp_NaK} shows the derived Na and K transmission spectra, which clearly hint a peak centered at Na and two peaks centered at the doublet of K. The derived Na and K transmission spectra are presented in the tables in Appendix \ref{sec:app_fig}.

From \citet{2015MNRAS.450..192W}'s tunable filter transit observations, we can obtain a radius difference of $\Delta R_\mathrm{p}/R_\star=0.0021\pm0.0018$ between 7664.9~\AA\ and 7582~\AA, which were observed simultaneously by switching between these two filters. Based on the DBSP 12~\AA\ bin at 7665~\AA\ and 50~\AA\ bin at 7582~\AA, we derived a radius difference of $\Delta R_\mathrm{p}/R_\star=0.0020\pm0.0022$, fully consistent with that of \citet{2015MNRAS.450..192W}. However, we note that \citet{2015MNRAS.450..192W}'s significant detection of K was based on the comparison between 7664.9~\AA\ and the weighted mean of 6792~\AA\ and 8844~\AA. Since the latter two were observed on very different dates with distinct systematics, we did not include them in this comparison.

%++++++++++++++++++++++++++++++++++
%   Table
%++++++++++++++++++++++++++++++++++
\begin{deluxetable*}{lcrrcrr}
\tablecaption{Parameter estimation for spectral retrievals assuming patchy clouds. \label{tab:ret_prior}}
\tablewidth{0pt}
\tablecolumns{7}
\tabletypesize{\small}
\tablehead{
\colhead{Parameter} & 
\colhead{Prior} & 
\multicolumn{2}{c}{Equilibrium chemistry} & 
\colhead{} & 
\multicolumn{2}{c}{Free chemistry}
\\
\cline{3-4}\cline{6-7}
\colhead{} & 
\colhead{} & 
\colhead{DBSP} & 
\colhead{DBSP+HST+Spitzer} & 
\colhead{} & 
\colhead{DBSP} & 
\colhead{DBSP+HST+Spitzer}
}
\startdata
$R_\mathrm{1bar}$(R$_\mathrm{J}$) & $\mathcal{U}(0.5,2)$ & $1.313^{+0.007}_{-0.009}$ & $1.298^{+0.006}_{-0.008}$ &
                                                         & $1.291^{+0.011}_{-0.012}$ & $1.256^{+0.009}_{-0.008}$\\
$T_\mathrm{iso}$(K) & $\mathcal{U}(500,1700)$ & $1082^{+175}_{-134}$ & $959^{+81}_{-40}$ &
                                              & $1562^{+95}_{-158}$  & $1643^{+40}_{-110}$\\
$\log P_\mathrm{cloud}$(bar) & $\mathcal{U}(-6,2)$ & $-1.5^{+2.2}_{-2.8}$ & $-1.0^{+2.1}_{-2.4}$ &
                                                   & $-1.7^{+2.3}_{-2.5}$ & $-1.8^{+2.4}_{-2.0}$\\
$\log A_\mathrm{RS}$ & $\mathcal{U}(0,10)$ & $3.2^{+3.8}_{-2.4}$ & $7.8^{+1.1}_{-1.5}$ &
                                           & $2.2^{+4.3}_{-1.6}$ & $8.3^{+1.0}_{-0.7}$\\
$\gamma$ & $\mathcal{U}(-20,2)$ & $-7.1^{+5.8}_{-7.6}$ & $-10.7^{+4.9}_{-5.8}$ &
                                & $-5.4^{+4.6}_{-7.0}$ & $-15.8^{+4.4}_{-2.7}$\\
$\phi$ & $\mathcal{U}(0,1)$ & $0.06^{+0.12}_{-0.05}$ & $0.21^{+0.07}_{-0.06}$ &
                            & $0.11^{+0.26}_{-0.08}$ & $0.22^{+0.05}_{-0.03}$\\
$\mathrm{C/O}$ & $\mathcal{U}(0.05,2)$ & $0.48^{+0.85}_{-0.29}$ & $0.25^{+0.23}_{-0.14}$ &
                                       & -- & --\\
$\log Z$(Z$_\sun$) & $\mathcal{U}(-1,3)$ & $0.52^{+0.45}_{-0.47}$ & $0.50^{+0.43}_{-0.55}$ &
                                         & -- & --\\
$\log X_\mathrm{Na}$   & $\mathcal{U}(-12,0)$ & -- & -- &
                                              & $-4.3^{+1.6}_{-1.0}$ & $-6.3^{+0.5}_{-0.4}$\\
$\log X_\mathrm{K}$    & $\mathcal{U}(-12,0)$ & -- & -- &
                                              & $-8.5^{+1.7}_{-1.8}$ & $-9.7^{+1.3}_{-1.4}$\\
$\log X_\mathrm{TiO}$  & $\mathcal{U}(-12,0)$ & -- & -- &
                                              & $-11.2^{+0.7}_{-0.5}$ & $-11.5^{+0.4}_{-0.3}$\\
$\log X_\mathrm{VO}$   & $\mathcal{U}(-12,0)$ & -- & -- &
                                              & $-10.9^{+0.9}_{-0.8}$ & $-11.1^{+0.7}_{-0.6}$\\
$\log X_\mathrm{H_2O}$ & $\mathcal{U}(-12,0)$ & -- & -- &
                                              & $-3.1^{+0.7}_{-0.8}$ & $-3.2^{+0.3}_{-0.3}$\\
$\log X_\mathrm{CH_4}$ & $\mathcal{U}(-12,0)$ & -- & -- &
                                              & $-6.9^{+3.5}_{-3.3}$ & $-8.5^{+2.4}_{-2.3}$\\
$\log X_\mathrm{CO}$   & $\mathcal{U}(-12,0)$ & -- & -- &
                                              & -- & $-7.6^{+2.9}_{-2.8}$\\
$\log X_\mathrm{CO_2}$ & $\mathcal{U}(-12,0)$ & -- & -- &
                                              & -- & $-9.1^{+1.9}_{-1.9}$\\
$\log X_\mathrm{HCN}$  & $\mathcal{U}(-12,0)$ & -- & -- &
                                              & -- & $-7.5^{+3.2}_{-2.9}$\\
$\log X_\mathrm{NH_3}$ & $\mathcal{U}(-12,0)$ & -- & -- &
                                              & -- & $-6.2^{+2.1}_{-3.8}$\\
\enddata
\end{deluxetable*}

%------------------------------------------------------------------------
%  5. ATMOSPHERIC SPECTRAL RETRIEVAL
%------------------------------------------------------------------------
\section{Atmospheric spectral retrieval}
\label{sec:retrieval}

We performed Bayesian spectral retrieval analyses on the transmission spectrum of HAT-P-1b to constrain its atmospheric properties. We employed the forward modeling module of the \texttt{PLATON} code \citep{2019PASP..131c4501Z,2020ApJ...899...27Z} and used \texttt{PyMultiNest} to explore the Bayesian evidence ($\mathcal{Z}$) and posterior distributions of the retrieval analyses. \texttt{PLATON} uses metallicity ($Z$) and the carbon-to-oxygen ratio (C/O) to parameterize the equilibrium chemistry model based on abundance grids computed by \texttt{GGchem} \citep{2018A&A...614A...1W}, which includes the same 34 atomic and molecular species as in \texttt{Exo-Transmit} \citep{2017PASP..129d4402K} but excluding C$_2$H$_6$ and SH. In our retrieval analyses, we adopted the opacities with a resolution of $\mathcal{R}=10000$ and assumed an isothermal temperature ($T_\mathrm{iso}$) for the atmosphere. We used the cloud-top pressure ($P_\mathrm{cloud}$) and the parametric form of $\sigma=A_\mathrm{RS}\sigma_0\lambda^{-\gamma}$ ($A_\mathrm{RS}$: an enhancement factor over nominal H$_2$ Rayleigh scattering) to account for clouds and hazes, which were assumed to occupy a fraction ($\phi$) of terminator atmosphere while the rest clear as in \citet{2017MNRAS.469.1979M}. The planet radius at a reference pressure of 1~bar ($R_\mathrm{1bar}$) was set as a free parameter. The planet mass was fixed to 0.525~M$_\mathrm{J}$ and the stellar radius was fixed to 1.174~R$_\sun$, taken from \citet{2014MNRAS.437...46N}.

We used 1000 live points in \texttt{PyMultiNest}, which typically resulted in an uncertainty of $\sim$0.1 for $\ln\mathcal{Z}$. We considered different model hypotheses in the following subsections. We adopted the criteria of \citet{2008ConPh..49...71T} to assess the model preference, where $\ln\mathcal{B}_{10}=\ln\mathcal{Z}_1-\ln\mathcal{Z}_0=1.0,2.5,5.0$ were interpreted as the starting points of ``weak'', ``moderate'', and ``strong'' evidence to prefer model hypothesis 1 over hypothesis 0. For the subsequent retrieval analyses, the derived parameters are listed in Table \ref{tab:ret_prior}, and their associated joint posterior distributions are presented in Appendix \ref{sec:app_fig_set2}. The statistics of the retrieval analyses are given in Table \ref{tab:ret_stat}. 

%++++++++++++++++++++++++++++++++++
%   Table
%++++++++++++++++++++++++++++++++++
\begin{deluxetable*}{lccccccccccc}
\tablecaption{Statistics from Bayesian spectral retrieval analyses. \label{tab:ret_stat}}
\tablewidth{0pt}
\tablecolumns{12}
\tabletypesize{\small}
\tablehead{
\colhead{Model} & 
\multicolumn{5}{c}{DBSP} &
\colhead{} & 
\multicolumn{5}{c}{DBSP+HST+Spitzer}
\\
\cline{2-6}\cline{8-12}
\colhead{} & 
\colhead{dof} & 
\colhead{$\chi^2_\mathrm{MAP}$\tablenotemark{\footnotesize a}} & 
\colhead{$\ln\mathcal{Z}$} & 
\colhead{$\Delta\ln\mathcal{Z}$} & 
\colhead{FS\tablenotemark{\footnotesize b}} &
\colhead{} & 
\colhead{dof} & 
\colhead{$\chi^2_\mathrm{MAP}$\tablenotemark{\footnotesize a}} & 
\colhead{$\ln\mathcal{Z}$} & 
\colhead{$\Delta\ln\mathcal{Z}$} & 
\colhead{FS\tablenotemark{\footnotesize b}}
}
\startdata
\multicolumn{12}{l}{\ \ 1. Wide coverage spectrum: equilibrium chemistry}\\
Patchy clouds   & 30 & 33.8 & 248.38 & 0 & Ref. & 
                & 72 & 85.4 & 550.99 & 0 & Ref.\\
Uniform clouds  & 31 & 32.3 & 245.88 & $-$2.5 & 2.8$\sigma$ & 
                & 73 & 93.0 & 544.80 & $-$6.2 & 3.9$\sigma$\\
\hline
\multicolumn{12}{l}{\ \ 2. Wide coverage spectrum: free chemistry w/ patchy clouds}\\
Full model & 26 & 23.2 & 247.30 & 0 & Ref. & 
           & 64 & 65.3 & 547.71 & 0 & Ref.\\
No Na      & 27 & 63.0 & 231.97 & $-$15.3 & 5.9$\sigma$ & 
           & 65 & 113.3 & 533.49 & $-$14.2 & 5.7$\sigma$\\
No K       & 27 & 23.8 & 247.09 & $-$0.2 & -- & 
           & 65 & 67.4 & 547.83 & 0.1 & --\\
No TiO     & 27 & 22.6 & 248.87 & 1.6 & $-$2.4$\sigma$ & 
           & 65 & 65.4 & 551.28 & 3.6 & $-$3.2$\sigma$\\
No VO      & 27 & 23.7 & 248.56 & 1.3 & $-$2.2$\sigma$ & 
           & 65 & 66.2 & 550.27 & 2.6 & $-$2.8$\sigma$\\
No H$_2$O  & 27 & 25.5 & 245.92 & $-$1.4 & 2.2$\sigma$ & 
           & 65 & 76.9 & 543.85 & $-$3.9 & 3.3$\sigma$\\
No CH$_4$  & 27 & 23.6 & 247.11 & $-$0.2 & -- & 
           & 65 & 66.6 & 548.67 & 1.0 & $-$2.0$\sigma$\\
No CO      & -- & -- & -- & -- & -- & 
           & 65 & 66.5 & 548.48 & 0.8 & --\\
No CO$_2$  & -- & -- & -- & -- & -- & 
           & 65 & 65.9 & 548.58 & 0.9 & --\\
No HCN     & -- & -- & -- & -- & -- & 
           & 65 & 66.1 & 548.34 & 0.6 & --\\
No NH$_3$  & -- & -- & -- & -- & -- & 
           & 65 & 69.5 & 547.40 & $-$0.3 & --\\
\hline
\multicolumn{12}{l}{\ \ 3. Na ultra-narrow band spectrum: free chemistry w/ uniform clouds}\\
Full model & 11 & 7.7  & 96.21 & 0 & Ref. & & -- & -- & -- & -- & --\\
No Na      & 12 & 26.8 & 89.25 & $-$7.0 &4.1$\sigma$ & & -- & -- & -- & -- & --\\
\hline
\multicolumn{12}{l}{\ \ 4. K ultra-narrow band spectrum: free chemistry w/ uniform clouds}\\
Full model & 7 & 6.9  & 72.15 & 0 & Ref. & & -- & -- & -- & -- & --\\
No K       & 8 & 22.3 & 66.55 & $-$5.6 &3.8$\sigma$ & & -- & -- & -- & -- & --\\
\enddata
\tablenotetext{a}{$\chi^2_\mathrm{MAP}$ stands for the $\chi^2$ of the maximum a posteriori (MAP) model.}
\tablenotetext{b}{Frequentist significance (FS) was converted from the Bayes factor $\mathcal{B}_{10}=\mathcal{Z}_1/\mathcal{Z}_0$ following \citet{2008ConPh..49...71T} and \citet{2013ApJ...778..153B}. FS is only calculated for those with $|\Delta\mathcal{Z}|\geq 1.0$. A positive FS value indicates that the reference model is favored over the model in comparison.}
\end{deluxetable*}

%------------------------------------------------------------------------
\subsection{DBSP transmission spectrum}
\label{sec:dbsp_transpec}

We first focused on the interpretation of our DBSP transmission spectrum. We started with the retrieval with the assumption of equilibrium chemistry. Only $R_\mathrm{1bar}$ and $\log Z$ show well constrained distributions, with the latter being $3.3^{+6.0}_{-2.2}\times$ solar metallicity. The other parameters are either unconstrained ($\log P_\mathrm{cloud}$, $\log A_\mathrm{RS}$, $\gamma$), bimodal ($T_\mathrm{iso}=1082^{+175}_{-134}$~K), or strongly skewed to the lower boundary ($\phi<0.27$, $\mathrm{C/O}<1.57$, 90\% upper limit). We find that the assumption of patchy clouds (free $\phi$) is moderately favored over uniform clouds ($\phi$ fixed to 1) by the DBSP spectrum given the latter has a reduced model evidence of $\Delta\ln\mathcal{Z}=-2.5$ (see Table \ref{tab:ret_stat}). However, both models can fit the DBSP spectrum at 1$\sigma$ confidence levels according to the $\chi^2$ statistic. 

\begin{figure}
\centering
\includegraphics[width=\columnwidth]{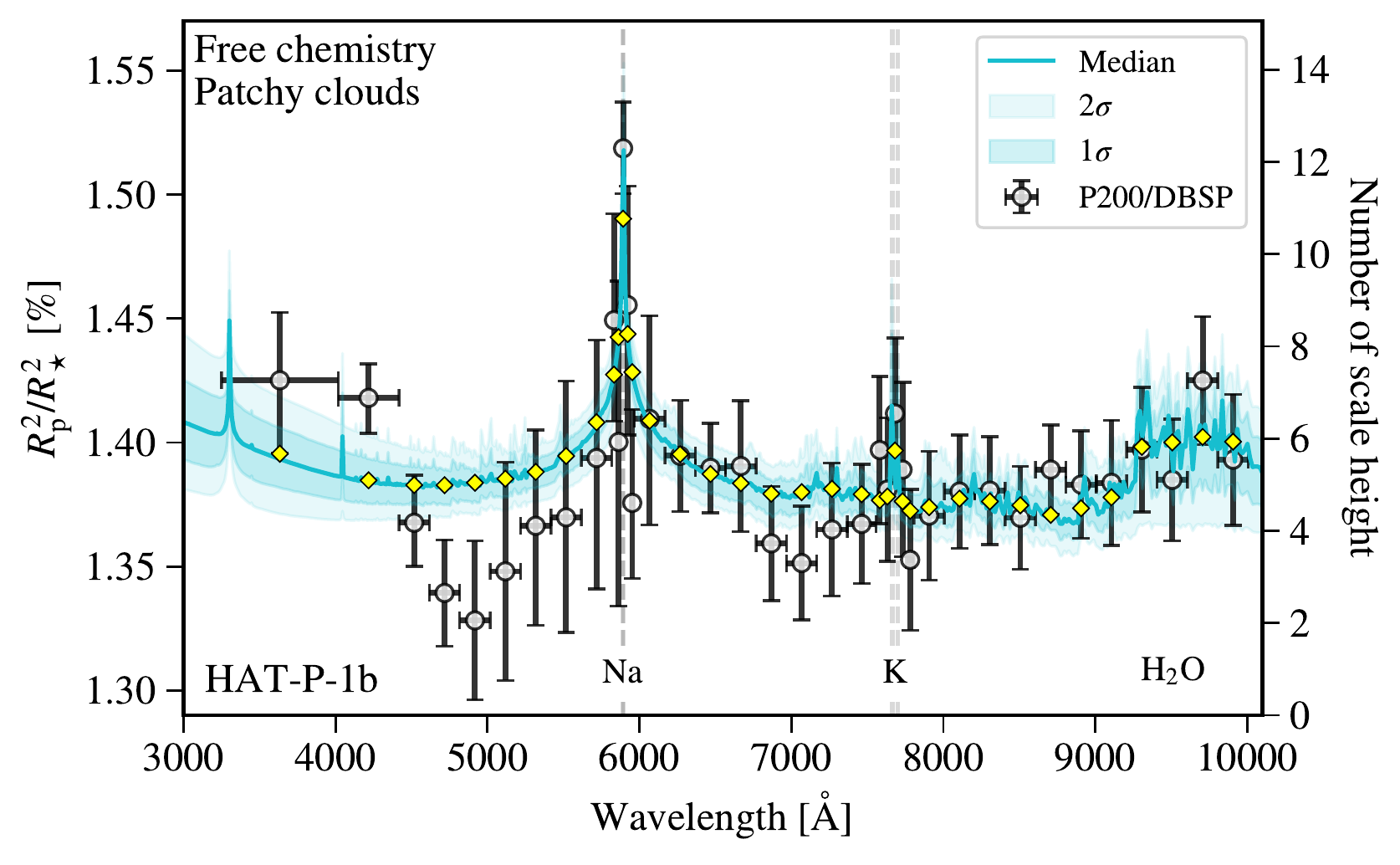}
\caption{Spectral retrieval analysis performed on the P200/DBSP data assuming free chemistry. The cyan line and shaded regions show the median and 1$\sigma$/2$\sigma$ confidence regions of the retrieved models.}
\label{fig:dbsp_retrieval}
\end{figure}

To explore the species responsible for the observed spectral signatures, we also performed the retrieval analysis with the assumption of free chemistry. In contrast to the equilibrium chemistry case, we only included the gas opacities of Na, K, TiO, VO, H$_2$O, and CH$_4$ in addition to H$_2$ and He, each of the former six had the volume mixing ratio ($X_i$) as a free parameter, which is constant across all pressures. Figure \ref{fig:dbsp_retrieval} shows the retrieved models assuming free chemistry. As presented in Table \ref{tab:ret_prior}, we obtained a much higher temperature of $T_\mathrm{iso}=1562^{+95}_{-158}$~K strongly skewed to the upper boundary ($>$1338~K, 90\% lower limit), while $\log P_\mathrm{cloud}$, $\log A_\mathrm{RS}$, and $\gamma$ remained unconstrained. The cloud coverage was still skewed to the lower boundary ($\phi<0.51$, 90\% upper limit). 

We then removed each of the six gas opacities one by one from the full model in additional retrievals. Such experiments reveal that the presence of Na is strongly favored because its absence would cause a significant decrease in model evidence ($\Delta\ln\mathcal{Z}=-15.3$; 5.9$\sigma$). The DBSP spectrum also moderately favors the presence of H$_2$O ($\Delta\ln\mathcal{Z}=-1.4$; 2.2$\sigma$). On the other hand, the presence of TiO or VO are disfavored at 2.4$\sigma$ and 2.2$\sigma$, respectively.

%------------------------------------------------------------------------
\subsubsection{Na and K ultra-narrow band transmission spectra}
We also performed the retrieval analysis on the Na and K ultra-narrow band transmission spectra individually. Due to fewer available data points compared to the full DBSP spectrum and much narrower wavelength coverage, we only included $T_\mathrm{iso}$, $R_\mathrm{1bar}$, $P_\mathrm{cloud}$, and $X_\mathrm{Na}$ (or $X_\mathrm{K}$) as free parameters, but fixed the other parameters ($\gamma=-4$, $\log A_\mathrm{RS}=0$, $\mathrm{C/O}=0.53$, $\log Z=0$, $\phi=1$). The prior on $T_\mathrm{iso}$ was still uniform, but the upper boundary was extended to 3000~K. Consequently, we derived $T_\mathrm{iso, Na}=2095^{+268}_{-340}$~K and $\log X_\mathrm{Na}=-4.7^{+2.2}_{-2.2}$ for the Na transmission spectrum, and $T_\mathrm{iso, K}=1725^{+392}_{-413}$~K and $\log X_\mathrm{K}=-5.3^{+2.6}_{-3.3}$ for the K transmission spectrum. As shown in Figure~\ref{fig:dbsp_NaK} and Table \ref{tab:ret_stat}, the retrieved models can fit the Na and K transmission spectra very well. 

We then muted the opacities of Na and K in an additional run of retrieval analysis, in order to inspect how the Na and K spectral signatures contribute to the models. Exclusion of Na and K would reduce the model evidence by $\Delta\ln\mathcal{Z}=-7.0$ for Na and $\Delta\ln\mathcal{Z}=-5.6$ for K, indicating that the presence of Na and K were both strongly favored in the ultra-narrow band transmission spectra. 

%++++++++++++++++++++++++++++++++++
%   Figure
%++++++++++++++++++++++++++++++++++

%++++++++++++++++++++++++++++++++++
%   Figure
%++++++++++++++++++++++++++++++++++
\begin{figure*}
\centering
\includegraphics[width=\textwidth]{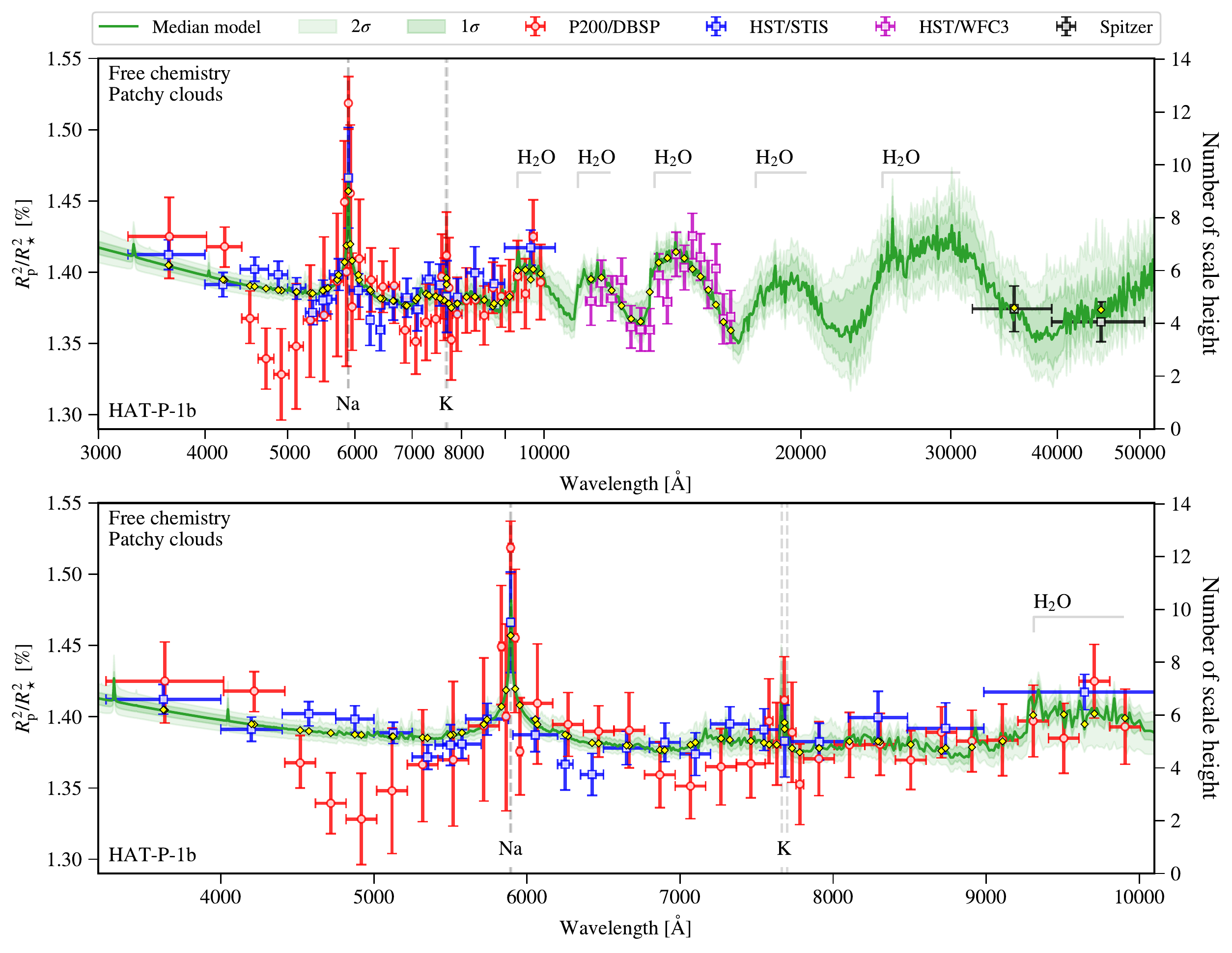}
\caption{{\it Top panel} presents the complete transmission spectrum of HAT-P-1b, with data from P200/DBSP, HST/STIS, HST/WFC3, and Spitzer. {\it Bottom panel} presents the close-up view of the optical wavelength range. The shaded areas show the retrieved atmospheric models assuming free chemistry. }
\label{fig:all_retrieved_freechem}
\end{figure*}

%------------------------------------------------------------------------
\subsection{Complete transmission spectrum}

For HAT-P-1b, high quality transmission spectrum has been observed from space by HST/STIS \citep{2014MNRAS.437...46N}, HST/WFC3 \citep{2013MNRAS.435.3481W}, and Spitzer \citep{2014MNRAS.437...46N}, covering a wavelength range from 3000 to 51000~\AA. Our DBSP spectrum has a wavelength coverage similar to that of HST/STIS, but has finer wavelength grids around the Na and K lines, thereby complementary to HST/STIS in corresponding wavelengths. We collected the revised HST and Spitzer data from \citet{2016Natur.529...59S}, and performed the equilibrium chemistry and free chemistry retrieval analyses on the DBSP, HST, and Spitzer combined dataset (hereafter complete spectrum), similar to that in Section~\ref{sec:dbsp_transpec}. A major difference is that more gas opacities were added in the free chemistry retrieval of the complete spectrum, including Na, K, TiO, VO, H$_2$O, CH$_4$, CO, CO$_2$, HCN, and NH$_3$. The retrieval results assuming free chemistry are shown in Figure~\ref{fig:all_retrieved_freechem}, while those assuming equilibrium chemistry can be found in Appendix \ref{sec:app_fig_set2}. 

The resulting parameters were in general consistent between the retrievals performed on the DBSP spectrum and the complete spectrum. For both equilibrium chemistry and free chemistry scenarios, the largest difference came from the 1~bar planet radius, which was smaller in the complete spectrum retrievals in both scenarios. The cloud-top pressure remained unconstrained. The scattering slope had a 90\% upper limit of $\gamma<-5.0$ in the equilibrium chemistry and $\gamma<-8.0$ in the free chemistry scenarios, respectively,  but the latter was strongly skewed to the lower boundary. The Rayleigh scattering enhancement factor and the cloud coverage were constrained to $A_\mathrm{RS}\sim 10^8$ and $\phi\sim0.22$ in both scenarios, respectively. For the equilibrium chemistry scenario, the derived metalicities were almost the same based on either the DBSP spectrum or the complete spectrum, while the C/O ratio was skewed to the lower boundary with a narrower distribution in the complete spectrum retrievals ($\mathrm{C/O}<0.54$, 90\% upper limit). On the other hand, the free chemistry scenario only resulted in well constrained volume mixing ratios for Na and H$_2$O, with smaller uncertainties ($\log X_\mathrm{Na}=-6.3^{+0.5}_{-0.4}$, $\log X_\mathrm{H_2O}=-3.2^{+0.3}_{-0.3}$) than those from the DBSP spectrum ($\log X_\mathrm{Na}=-4.3^{+1.6}_{-1.0}$, $\log X_\mathrm{H_2O}=-3.1^{+0.7}_{-0.8}$). 

Similar to the results from the DBSP spectrum, the equilibrium chemistry retrieval of the complete spectrum favored the assumption of patchy clouds at 3.9$\sigma$, while the free chemistry retrieval favored the presence of Na (5.7$\sigma$) and H$_2$O (3.3$\sigma$), but disfavored TiO (3.2$\sigma$), VO (2.8$\sigma$), and CH$_4$ (2.0$\sigma$). Therefore, the DBSP spectrum delivered very consistent information with that of the 3000--51000~{\AA} complete spectrum, with larger uncertainties though.

%------------------------------------------------------------------------
%  6. DISCUSSION
%------------------------------------------------------------------------
\section{Discussion}
\label{sec:discuss}

%------------------------------------------------------------------------
\subsection{Comparison with literature results}
Our new DBSP observations of HAT-P-1b have revealed a broad spectral signature around the Na line at 5.9$\sigma$, indicating the presence of a pressure broadened line wing for Na. The reasonable precision and spectral resolution of DBSP allow the creation of uniformly spaced 12~\AA\ transmission spectra centered at Na and K, resulting in the detection of the line cores of both Na (4.1$\sigma$) and K (3.8$\sigma$) doublets. 

To search for excess absorption at the Na and K lines, the approach of comparing a central narrow band to the neighboring wide bands has been widely adopted in both space- and ground-based transit spectrophotometry due to limited narrow-band precision and spectral resolution \citep[e.g.,][]{2014MNRAS.437...46N}. An alternative way is to perform quasi-simultaneous narrow-band photometry by switching between two tunable filters designed to cover the core and wing of a spectral line \citep[e.g.,][]{2015MNRAS.450..192W}. However, both approaches have the risk of obtaining false positive detections due to noise fluctuation in only one or two narrow bands. 

In contrast, tens of uniform narrow bands centered around the target line could in principal perform better in retaining the noise fluctuation and potential excess absorption signal. The latter has been successfully implemented to detect Na in several exolplanets \citep[e.g.,][]{2011A&A...527A..73S,2017A&A...600L..11C,2018A&A...616A.145C,2020A&A...642A..54C,2018Natur.557..526N,2019AJ....157...21P,2022MNRAS.510.4857A}. Our detection of Na and K in HAT-P-1b is another example. 

Several studies have performed spectral retrieval analyses on HAT-P-1b's HST and Spitzer combined dataset \citep{2017ApJ...834...50B,2019MNRAS.482.1485P,2019ApJ...887L..20W}. Our retrieval approach approximately follows \citet{2019MNRAS.482.1485P} and \citet{2019ApJ...887L..20W}, in particular the assumption of cloudy and clear sectors. The latter two have obtained almost the same retrieval results. \citet{2019ApJ...887L..20W} reported the detection of H$_2$O at 3.2$\sigma$ ($\log X_\mathrm{H_2O}=-2.5^{+0.8}_{-0.7}$) and marginal evidence of Na at 1.3$\sigma$ ($\log X_\mathrm{Na}=-8.6^{+1.2}_{-1.8}$). With the addition of the newly acquired DBSP data, we are able to detect H$_2$O at 3.3$\sigma$ ($\log X_\mathrm{H_2O}=-3.2^{+0.3}_{-0.3}$) and Na at 5.7$\sigma$ ($\log X_\mathrm{Na}=-6.3^{+0.5}_{-0.4}$). The retrieval performed on the DBSP data alone results in a H$_2$O volume mixing ratio ($\log X_\mathrm{H_2O}=-3.1^{+0.7}_{-0.8}$), almost the same as that from the DBSP, HST, and Spitzer combined dataset, revealing a weak evidence for the 9300~{\AA} H$_2$O feature. On the other hand, we derive a well constrained cloud fraction of $0.22^{+0.05}_{-0.03}$, much smaller than $\sim$$0.5^{+0.1}_{-0.1}$ reported in \citet{2019MNRAS.482.1485P} and \citet{2019ApJ...887L..20W}.

%------------------------------------------------------------------------
\subsection{Interpreting the atmosphere of HAT-P-1b}
Our retrieval results suggest that HAT-P-1b's terminator atmosphere is mostly clear, with a small fraction of $\sim$22\% covered by clouds and hazes. The inference of a large clear region is probably due to the presence of the pressure broadened line wing of Na and prominent spectral signature of H$_2$O. However, we are not able to detect the pressure broadening of the K line. This is not a rare phenomenon in hot Jupiters. Presence of Na pressure broadening and absence of K pressure broadening have been also found in WASP-96b \citep{2018Natur.557..526N,2022MNRAS.tmp.1485N}, WASP-21b \citep{2020A&A...642A..54C}, and WASP-62b \citep{2021ApJ...906L..10A}. 

In our retrieved cloudy and hazy region, haze is the dominant scattering opacity source. \citet{2022MNRAS.tmp.1720A} recently studied the photochemical hazes in the atmospheres of ten hot Jupiters based on a self-consistent model including haze microphysics, disequilibrium chemistry, and radiative feedback. HAT-P-1b is among the planets showing degeneracies, for which the low-haze or haze-free solution was presented in their discussion. According to their neglected hazy solution, HAT-P-1b would have a haze density of 8~particle\,cm$^{-3}$ for a mass flux of 10$^{-12}$~g\,cm$^{-2}$\,s$^{-1}$. This might be the case for the cloudy and hazy region as retrieved by our analyses.

In the equilibrium chemistry retrieval of the DBSP, HST, and Spitzer combined dataset, we obtained a low C/O of 0.25$^{+0.23}_{-0.14}$ and a metallicity of 3.2$^{+5.3}_{-2.3}$$\times$ solar metallicity. Our retrieved metallicity is fully consistent with the value of 3.6$^{+5.8}_{-2.6}$ derived by \citet{2019MNRAS.482.1485P}.  The low C/O value is a natural result of non-detection of carbon-baring molecules as evidenced by the free chemistry retrieval. The latter has resulted in a better goodness of fit than the equilibrium chemistry retrieval according to the reduced $\chi^2_\mathrm{MAP}$. The non-detection might be caused by the lack of resolvable carbon-related spectral signatures in the current wavelength coverage, where H$_2$O is the dominant opacity source. The Spitzer data could in principal provide constraints on carbon-baring molecules, but could be easily biased by systematics when only two very broad bands are available. 

We calculate the C/H, O/H, Na/H, K/H ratios relative to solar values \citep{2009ARA&A..47..481A} in logarithmic scale from the free chemistry retrieval results, and obtain $\mathrm{[C/H]}<-0.18$ (90\% upper limit), $\mathrm{[O/H]}=-0.15^{+0.32}_{-0.28}$, $\mathrm{[Na/H]}=-0.77^{+0.50}_{-0.40}$, and $\mathrm{[K/H]}<-1.25$ (90\% upper limit), of which only the O and Na abundances are constrained. All the abundances appear sub-solar, except for O being consistent with the solar value within uncertainties. As the major O carrier, the abundance of H$_2$O is consistent with the solar value of $-3.3$ \citep{2009ARA&A..47..481A,2012ApJ...758...36M,2019MNRAS.482.1485P}. When compared to that of the host star, i.e., $\mathrm{[C/H]}=-0.001\pm  0.031$, $\mathrm{[O/H]}=+0.088\pm 0.035$, and $\mathrm{[Na/H]}=+0.090\pm 0.011$ \citep{2014MNRAS.442L..51L}, the planetary abundances are in the range of substellar to stellar. 

With a low C/O ratio ($<$0.54, 90\% upper limit), a low C/H ratio ($<$0.7$\times$solar, 90\% upper limit), and a solar O/H ratio (0.4--1.5$\times$solar, 68.3\% confidence interval), HAT-P-1b could be formed through either core accretion with disk migration or gravitational instability with disk-free migration according to the formation-migration pathways presented by \citet{2014ApJ...794L..12M}. On the other hand, HAT-P-1b has a refractory-to-oxygen ratio of $\mathrm{[R/O]}=-0.67^{+0.52}_{-0.45}$ based on Na, K, TiO, and VO from our free chemistry retrieval, which favors an icy instead of rocky-rich planetesimal enrichment \citep{2021ApJ...914...12L}. However, the R/O ratio is proposed for ultrahot Jupiters, while HAT-P-1b's equilibrium temperature is much cooler than them. A large fraction of the refractory elements could have been condensed into solids, thus absence from observable atmosphere.

%------------------------------------------------------------------------
%  7. CONCLUSIONS
%------------------------------------------------------------------------
\section{Conclusions}
\label{sec:concl}

We have presented a new optical transmission spectrum of HAT-P-1b based on two transits observed with P200/DBSP. The DBSP transmission spectrum exhibits a pressure broadened line wing at the Na line and a tentative H$_2$O absorption signature at around 9300~\AA. The line cores of Na and K are spectrally resolved when the passband bin width is as narrow as 12~\AA. Our spectral retrieval analyses on the DBSP spectrum reveal a mostly clear atmosphere for HAT-P-1b, with the presence of Na line wing detected at 5.9$\sigma$, Na line core at 4.1$\sigma$, K line core at 3.8$\sigma$, and H$_2$O at 2.2$\sigma$. The results from the retrievals of the DBSP, HST/STIS, HST/WFC3, and Spitzer combined dataset are consistent with that performed on the DBSP data alone, but are generally more precise and the significance of H$_2$O increases to 3.3$\sigma$. The current available dataset suggests a metallicity of 3.2$^{+5.3}_{-2.3}$$\times$ solar, a C/O ratio lower than 0.54 at 90\% confidence level, subsolar values for Na, K, C, and a subsolar-to-solar value for O. 

Future high-resolution Doppler spectroscopy could resolve Na and K and provide insight into the atmospheric dynamics in the upper atmosphere of HAT-P-1b. The JWST GTO-1201 program and any future JWST observations should be able to put stringent constraints on all the important oxygen, carbon, and nitrogen carriers, enabling an in-depth exploration of HAT-P-1b's formation and migration history.

\begin{acknowledgments}
\vspace{0.1em}% this line is not necessary if the layout looks ok.

G.\,C. acknowledges the support by the National Natural Science Foundation of China (Grant No. 42075122, 12122308), the Natural Science Foundation of Jiangsu Province (Grant No.\,BK20190110), Youth Innovation Promotion Association CAS (2021315), and the Minor Planet Foundation of the Purple Mountain Observatory. The authors would like to thank Carolyn Heffner and Kajsa Peffer for their great supports during the observations, Sheng Jin for carrying out the second observation, Nikku Madhusudhan and Neale Gibson for their help at the early stage of this work. This research uses data obtained through the Telescope Access Program (TAP), which has been funded by the TAP member institutes. Observations obtained with the Hale Telescope at Palomar Observatory were obtained as part of an agreement between the National Astronomical Observatories, Chinese Academy of Sciences, and the California Institute of Technology. We acknowledge the use of TESS High Level Science Products (HLSP) produced by the Quick-Look Pipeline (QLP) at the TESS Science Office at MIT, which are publicly available from the Mikulski Archive for Space Telescopes (MAST). Funding for the TESS mission is provided by NASA's Science Mission directorate.
\end{acknowledgments}


\vspace{5mm}
\facilities{Palomar 200-inch (DBSP)}


\software{
\texttt{george} \citep{2015ITPAM..38..252A}, 
\texttt{batman} \citep{2015PASP..127.1161K},
\texttt{emcee} \citep{2013PASP..125..306F}, 
\texttt{corner} \citep{2016JOSS....1...24F}, 
\texttt{PLATON} \citep{2019PASP..131c4501Z,2020ApJ...899...27Z}, 
\texttt{PyMultiNest} \citep{2014A&A...564A.125B}, 
\texttt{Matplotlib} \citep{2007CSE.....9...90H}, 
\texttt{TEPCat} \citep{2011MNRAS.417.2166S},
\texttt{VizieR} \citep{2000A&AS..143...23O}
}



%------------------------------------------------------------------------
%  APPENDIX
%------------------------------------------------------------------------
\appendix

%------------------------------------------------------------------------
\section{Spectroscopic light curves}
\label{sec:app_fig}

\begin{figure}
\centering
\includegraphics[width=\textwidth]{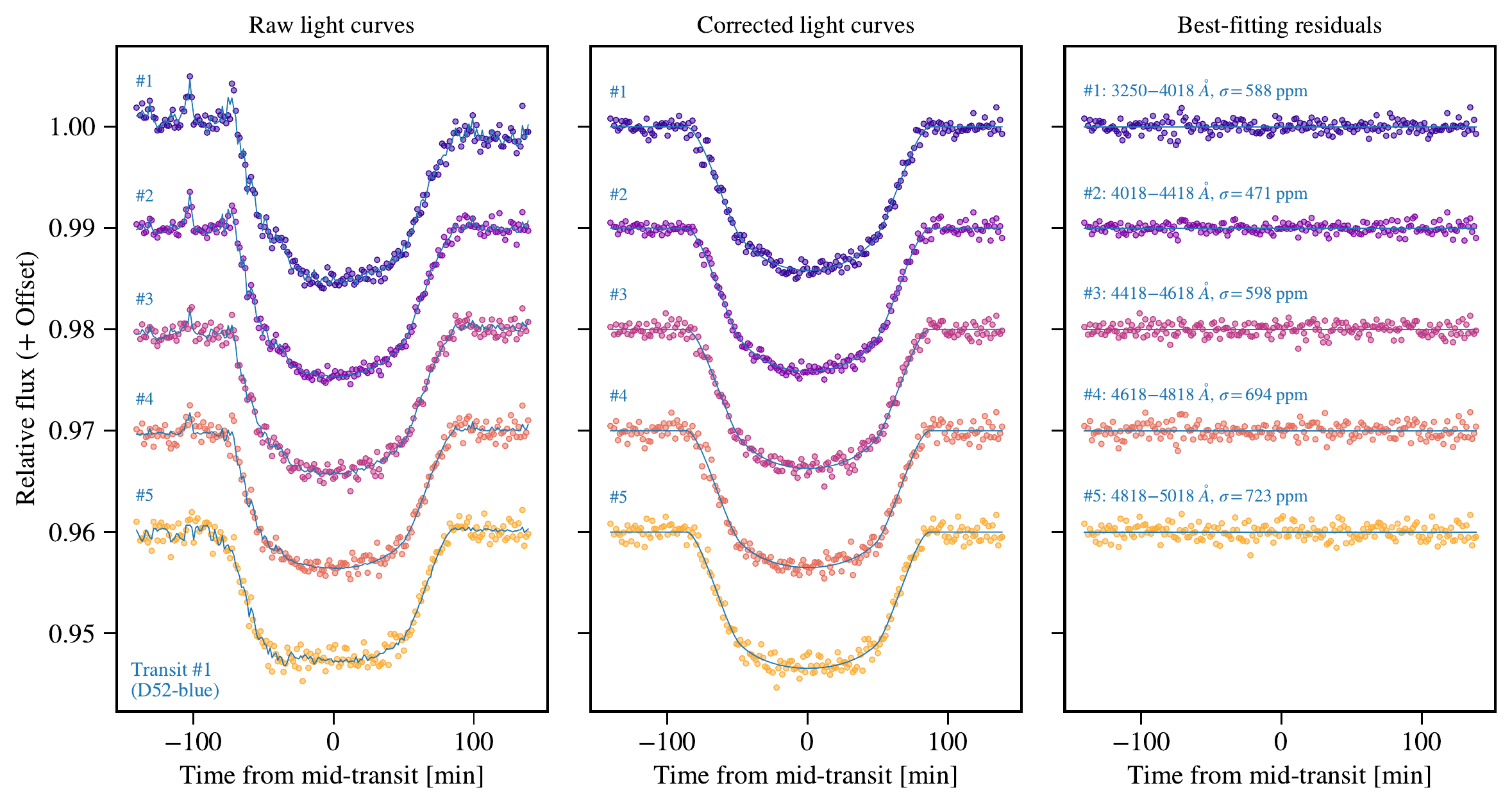}
\caption{Spectroscopic light curves in the D52-blue channel obtained with P200/DBSP on the night of 2012 September 14. From left to right presents the reference-calibrated raw light curves, systematics corrected light curves, and best-fitting residuals, respectively.}
\label{fig:dbsp_slc_d52b}
\end{figure}

\begin{figure}
\centering
\includegraphics[width=\textwidth]{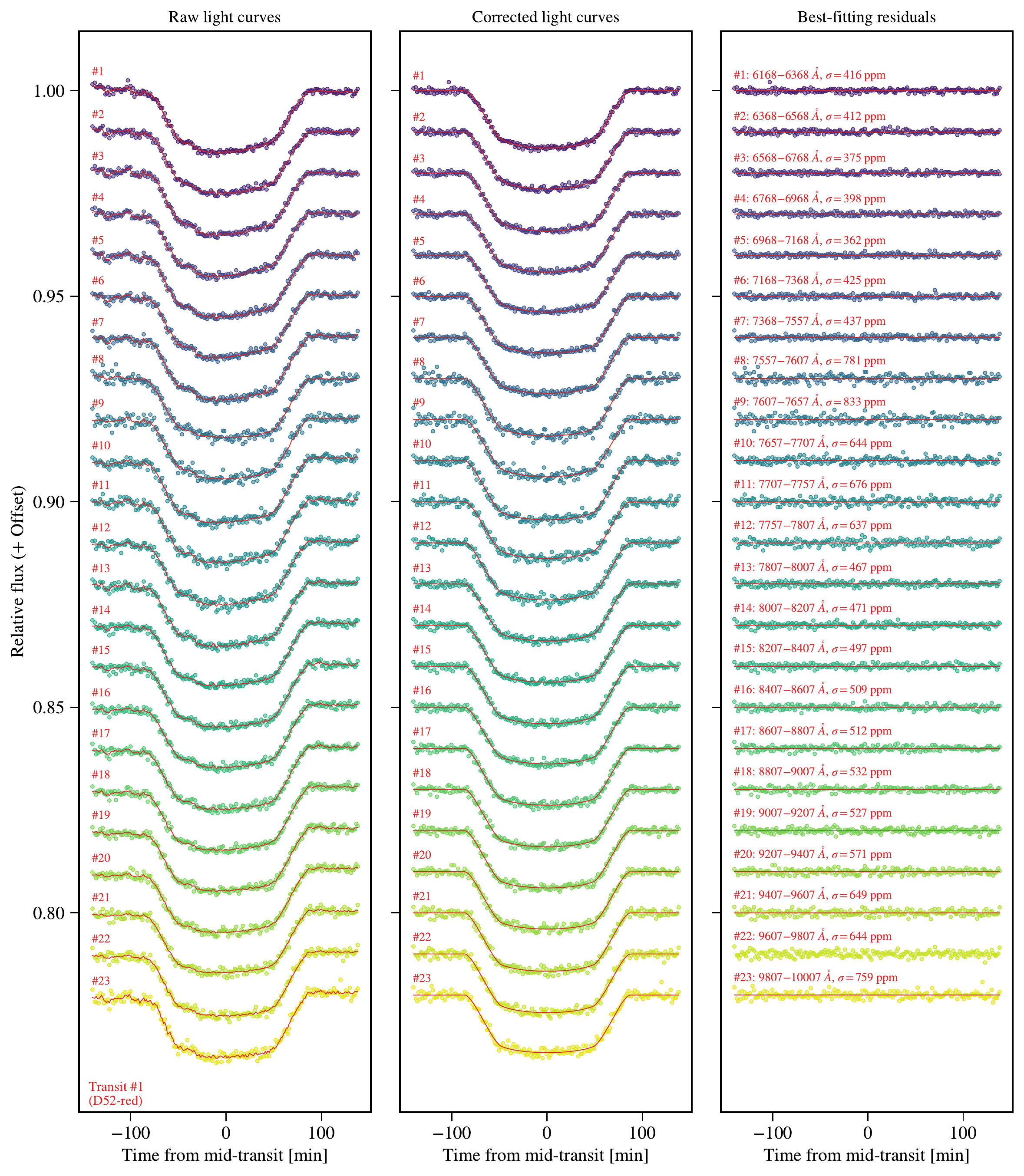}
\caption{Same as Figure~\ref{fig:dbsp_slc_d52b}, but for the D52-red channel.}
\label{fig:dbsp_slc_d52r}
\end{figure}

\begin{figure}
\centering
\includegraphics[width=\textwidth]{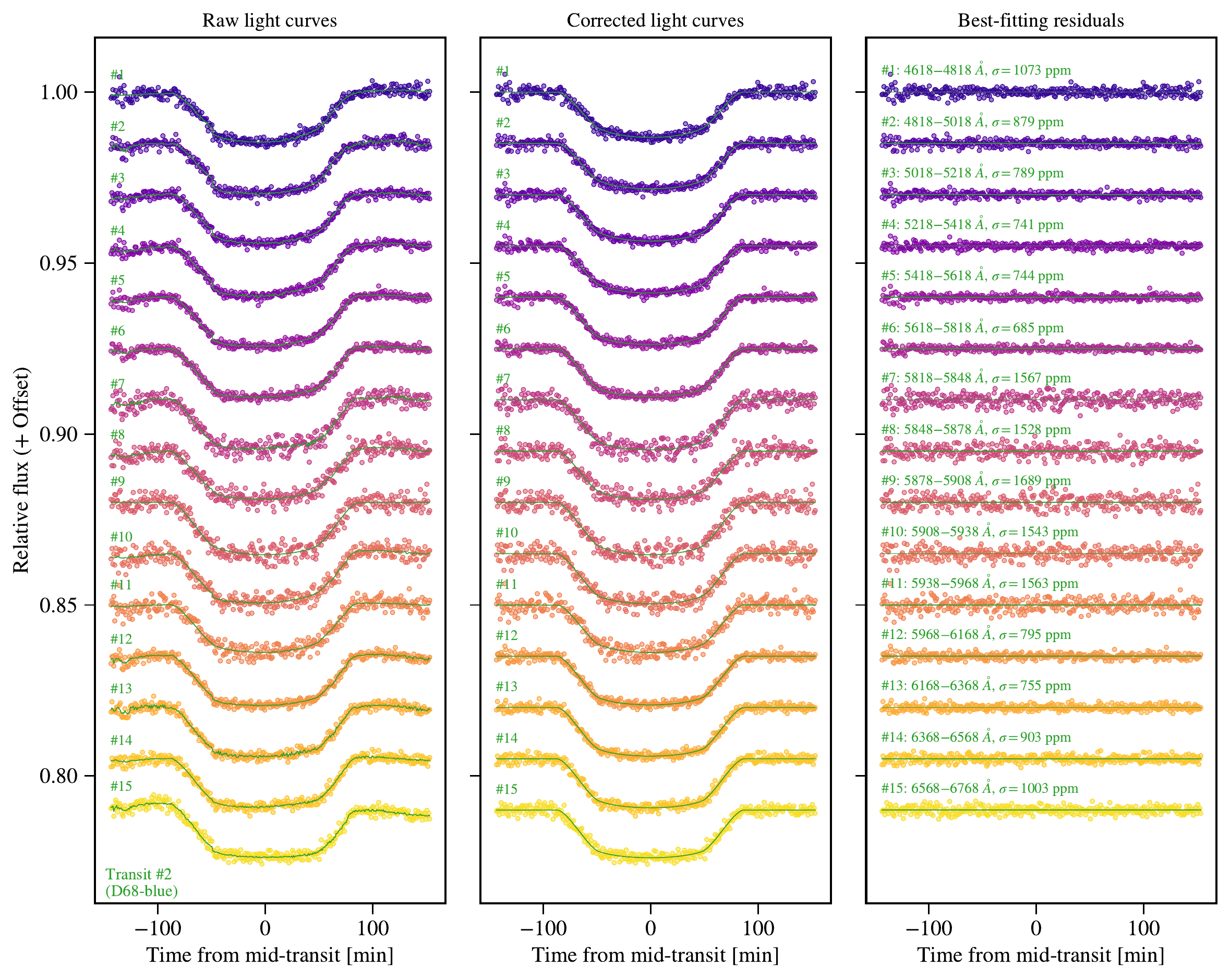}
\caption{Spectroscopic light curves in the D68-blue channel obtained with P200/DBSP on the night of 2016 July 26. From left to right presents the reference-calibrated raw light curves, systematics corrected light curves, and best-fitting residuals, respectively.}
\label{fig:dbsp_slc_d68b}
\end{figure}

\begin{figure}
\centering
\includegraphics[width=\textwidth]{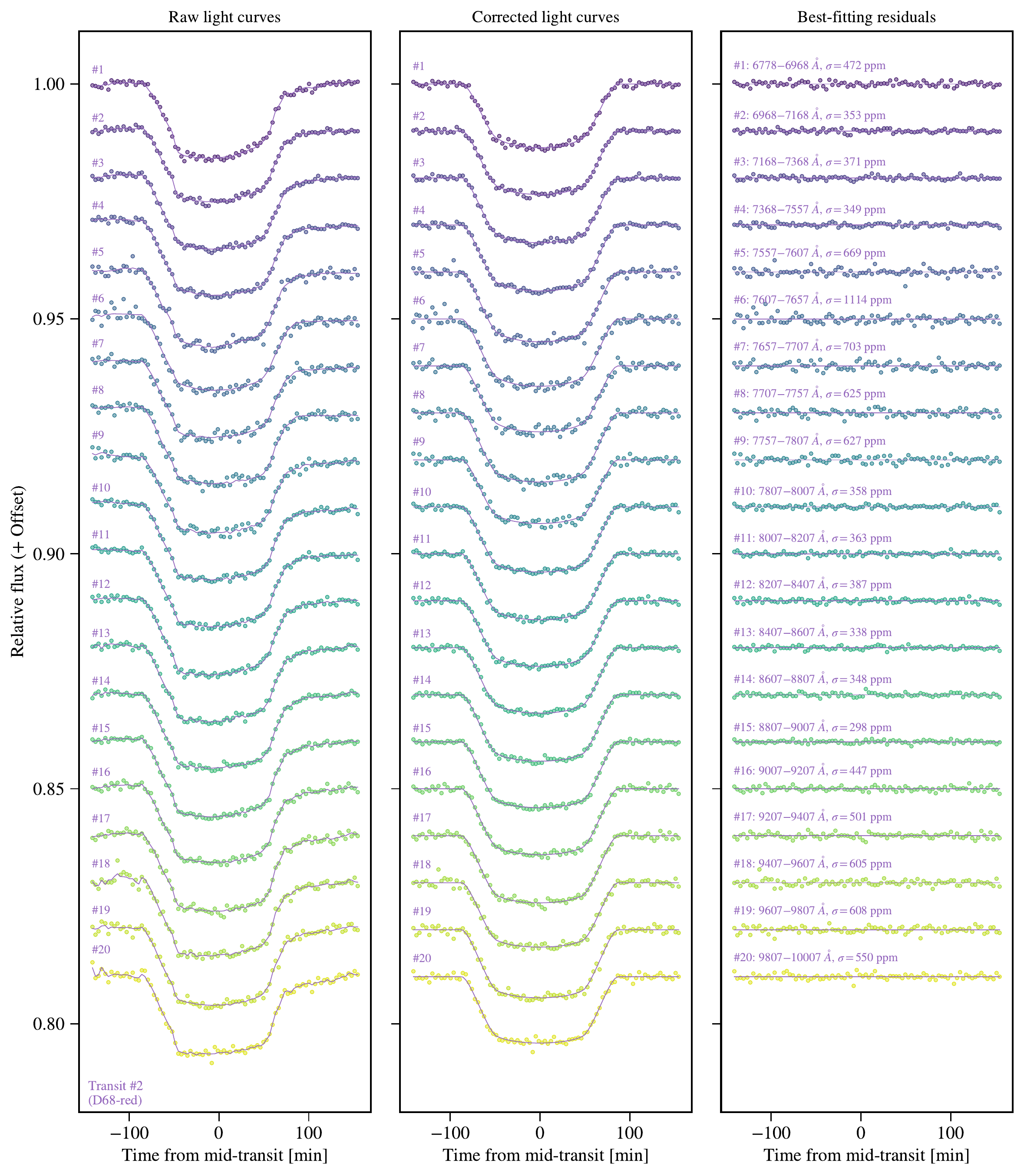}
\caption{Same as Figure~\ref{fig:dbsp_slc_d68b}, but for the D68-red channel.}
\label{fig:dbsp_slc_d68r}
\end{figure}

The raw spectroscopic light curves, along with the systematics corrected ones and the best-fit residuals, are presented in Figure \ref{fig:dbsp_slc_d52b} for D52-blue, Figure \ref{fig:dbsp_slc_d52r} for D52-red, Figure \ref{fig:dbsp_slc_d68b} for D68-blue, and Figure \ref{fig:dbsp_slc_d68r} for D68-red, respectively. The derived Na and K ultra-narrow band transmission spectra are presented in Table \ref{tab:ts_nak}.

%++++++++++++++++++++++++++++++++++
%   Table
%++++++++++++++++++++++++++++++++++
\begin{deluxetable}{cccc}
\tablecaption{Ultra-narrow band transmission spectra of HAT-P-1b.\label{tab:ts_nak}}
\tablewidth{0pt}
\tabletypesize{\small}
\tablehead{
\colhead{$\lambda$ (\AA)} & 
\colhead{$u_1$ prior} & 
\colhead{$u_2$ prior} &
\colhead{$R_\mathrm{p}/R_\star$}
} 
\startdata
\multicolumn{4}{c}{Na transmission spectrum\tablenotemark{\footnotesize a}}\\
    5803--5815 & $\mathcal{N}(0.395,0.045^2)$ & $\mathcal{N}(0.294,0.025^2)$ & $0.1179 ^{+0.0010}_{-0.0010}$\\
    5815--5827 & $\mathcal{N}(0.391,0.046^2)$ & $\mathcal{N}(0.298,0.026^2)$ & $0.1172 ^{+0.0010}_{-0.0010}$\\
    5827--5839 & $\mathcal{N}(0.388,0.046^2)$ & $\mathcal{N}(0.300,0.026^2)$ & $0.1192 ^{+0.0009}_{-0.0010}$\\
    5839--5851 & $\mathcal{N}(0.389,0.045^2)$ & $\mathcal{N}(0.297,0.025^2)$ & $0.1195 ^{+0.0016}_{-0.0021}$\\
    5851--5863 & $\mathcal{N}(0.388,0.045^2)$ & $\mathcal{N}(0.297,0.025^2)$ & $0.1188 ^{+0.0012}_{-0.0013}$\\
    5863--5875 & $\mathcal{N}(0.386,0.046^2)$ & $\mathcal{N}(0.298,0.026^2)$ & $0.1188 ^{+0.0010}_{-0.0010}$\\
    5875--5887 & $\mathcal{N}(0.391,0.046^2)$ & $\mathcal{N}(0.288,0.028^2)$ & $0.1219 ^{+0.0034}_{-0.0032}$\\
    5887--5899 & $\mathcal{N}(0.394,0.047^2)$ & $\mathcal{N}(0.282,0.029^2)$ & $0.1242 ^{+0.0031}_{-0.0031}$\\
    5899--5911 & $\mathcal{N}(0.387,0.046^2)$ & $\mathcal{N}(0.292,0.027^2)$ & $0.1210 ^{+0.0010}_{-0.0010}$\\
    5911--5923 & $\mathcal{N}(0.383,0.045^2)$ & $\mathcal{N}(0.297,0.025^2)$ & $0.1211 ^{+0.0010}_{-0.0010}$\\
    5923--5935 & $\mathcal{N}(0.381,0.045^2)$ & $\mathcal{N}(0.298,0.025^2)$ & $0.1189 ^{+0.0010}_{-0.0010}$\\
    5935--5947 & $\mathcal{N}(0.380,0.045^2)$ & $\mathcal{N}(0.298,0.025^2)$ & $0.1179 ^{+0.0011}_{-0.0010}$\\
    5947--5959 & $\mathcal{N}(0.379,0.045^2)$ & $\mathcal{N}(0.299,0.025^2)$ & $0.1175 ^{+0.0010}_{-0.0010}$\\
    5959--5971 & $\mathcal{N}(0.376,0.044^2)$ & $\mathcal{N}(0.300,0.024^2)$ & $0.1179 ^{+0.0010}_{-0.0010}$\\
    5971--5983 & $\mathcal{N}(0.375,0.044^2)$ & $\mathcal{N}(0.301,0.024^2)$ & $0.1160 ^{+0.0011}_{-0.0010}$\\
\hline
\multicolumn{4}{c}{K transmission spectrum\tablenotemark{\footnotesize b}}\\
    7623--7635 & $\mathcal{N}(0.275,0.035^2)$ & $\mathcal{N}(0.294,0.017^2)$ & $0.1160 ^{+0.0009}_{-0.0009}$\\
    7635--7647 & $\mathcal{N}(0.275,0.034^2)$ & $\mathcal{N}(0.293,0.016^2)$ & $0.1165 ^{+0.0007}_{-0.0007}$\\
    7647--7659 & $\mathcal{N}(0.274,0.033^2)$ & $\mathcal{N}(0.292,0.015^2)$ & $0.1169 ^{+0.0021}_{-0.0020}$\\
    7659--7671 & $\mathcal{N}(0.273,0.033^2)$ & $\mathcal{N}(0.290,0.015^2)$ & $0.1195 ^{+0.0017}_{-0.0017}$\\
    7671--7682 & $\mathcal{N}(0.272,0.033^2)$ & $\mathcal{N}(0.291,0.016^2)$ & $0.1188 ^{+0.0013}_{-0.0013}$\\
    7682--7693 & $\mathcal{N}(0.272,0.034^2)$ & $\mathcal{N}(0.292,0.016^2)$ & $0.1184 ^{+0.0007}_{-0.0007}$\\
    7693--7705 & $\mathcal{N}(0.271,0.034^2)$ & $\mathcal{N}(0.293,0.016^2)$ & $0.1193 ^{+0.0006}_{-0.0006}$\\
    7705--7717 & $\mathcal{N}(0.271,0.034^2)$ & $\mathcal{N}(0.292,0.016^2)$ & $0.1174 ^{+0.0009}_{-0.0014}$\\
    7717--7729 & $\mathcal{N}(0.273,0.034^2)$ & $\mathcal{N}(0.291,0.016^2)$ & $0.1183 ^{+0.0007}_{-0.0007}$\\
    7729--7741 & $\mathcal{N}(0.273,0.034^2)$ & $\mathcal{N}(0.290,0.016^2)$ & $0.1169 ^{+0.0006}_{-0.0007}$\\
    7741--7753 & $\mathcal{N}(0.270,0.034^2)$ & $\mathcal{N}(0.290,0.016^2)$ & $0.1167 ^{+0.0006}_{-0.0007}$\\
\enddata
\tablenotetext{a}{The Na transmission spectrum was derived from common-mode corrected spectroscopic light curves, and the common-mode systematics was derived from the 5800-6000~\AA\ band, with $R_\mathrm{p}/R_\star=0.1193\pm0.0014$.}
\tablenotetext{b}{The K transmission spectrum was derived from common-mode corrected spectroscopic light curves, and the common-mode systematics was derived from the 7600-7800~\AA\ band, with $R_\mathrm{p}/R_\star=0.1176\pm0.0010$.}
\end{deluxetable}

%++++++++++++++++++++++++++++++++++
%   Figure
%++++++++++++++++++++++++++++++++++

%++++++++++++++++++++++++++++++++++
%   Figure
%++++++++++++++++++++++++++++++++++

%++++++++++++++++++++++++++++++++++
%   Figure
%++++++++++++++++++++++++++++++++++

%++++++++++++++++++++++++++++++++++
%   Figure
%++++++++++++++++++++++++++++++++++

%------------------------------------------------------------------------
\section{Additional plots from the Bayesian spectral retrieval}
\label{sec:app_fig_set2}

\begin{figure}
\centering
\includegraphics[width=\textwidth]{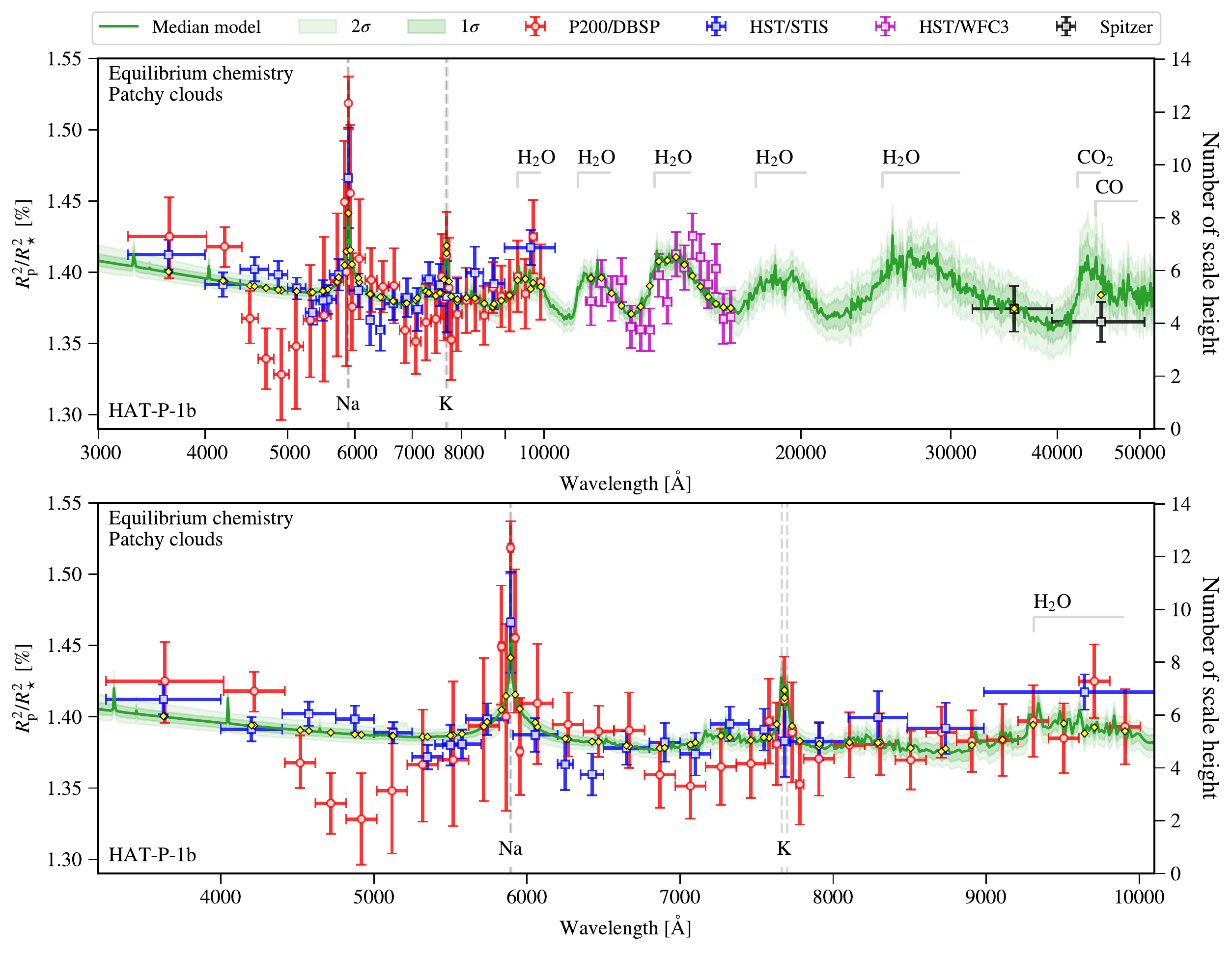}
\caption{{\it Top panel} presents the complete transmission spectrum of HAT-P-1b, with data from P200/DBSP, HST/STIS, HST/WFC3, and Spitzer. {\it Bottom panel} presents the close-up view of the optical wavelength range. The shaded areas show the retrieved atmospheric models assuming equilibrium chemistry. }
\label{fig:all_retrieved_eqchem}
\end{figure}

The retrieved transmission spectra assuming equilibrium chemistry are shown in Figure \ref{fig:all_retrieved_eqchem}, which is in a similar manner to the free chemistry retrievals as shown in Figure \ref{fig:all_retrieved_freechem}. 

\begin{figure}
\centering
\includegraphics[width=\textwidth]{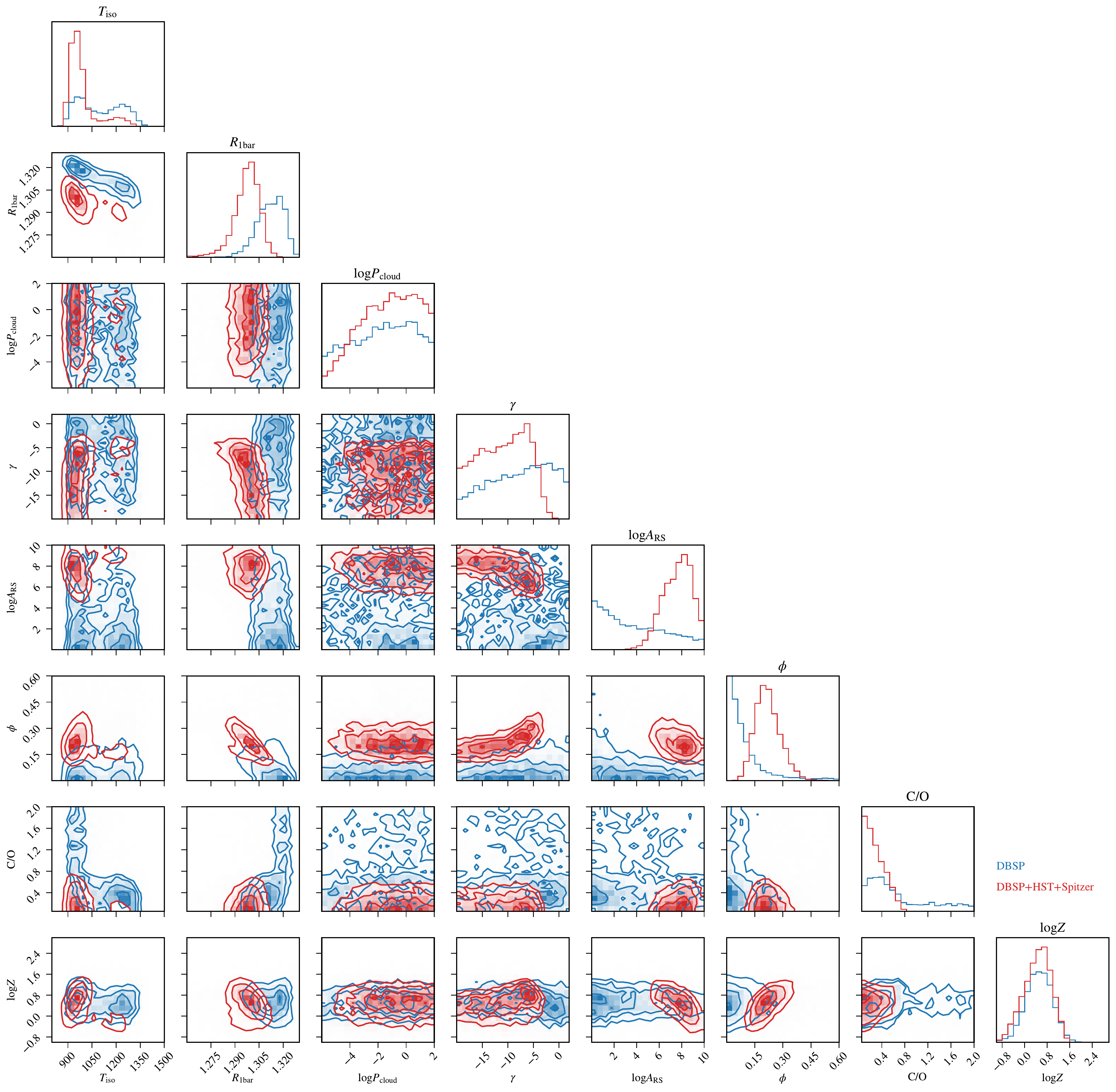}
\caption{Corner plot for spectral retrieval analyses assuming equilibrium chemistry.}
\label{fig:corner_eqchem}
\end{figure}

\begin{figure}
\centering
\includegraphics[width=\textwidth]{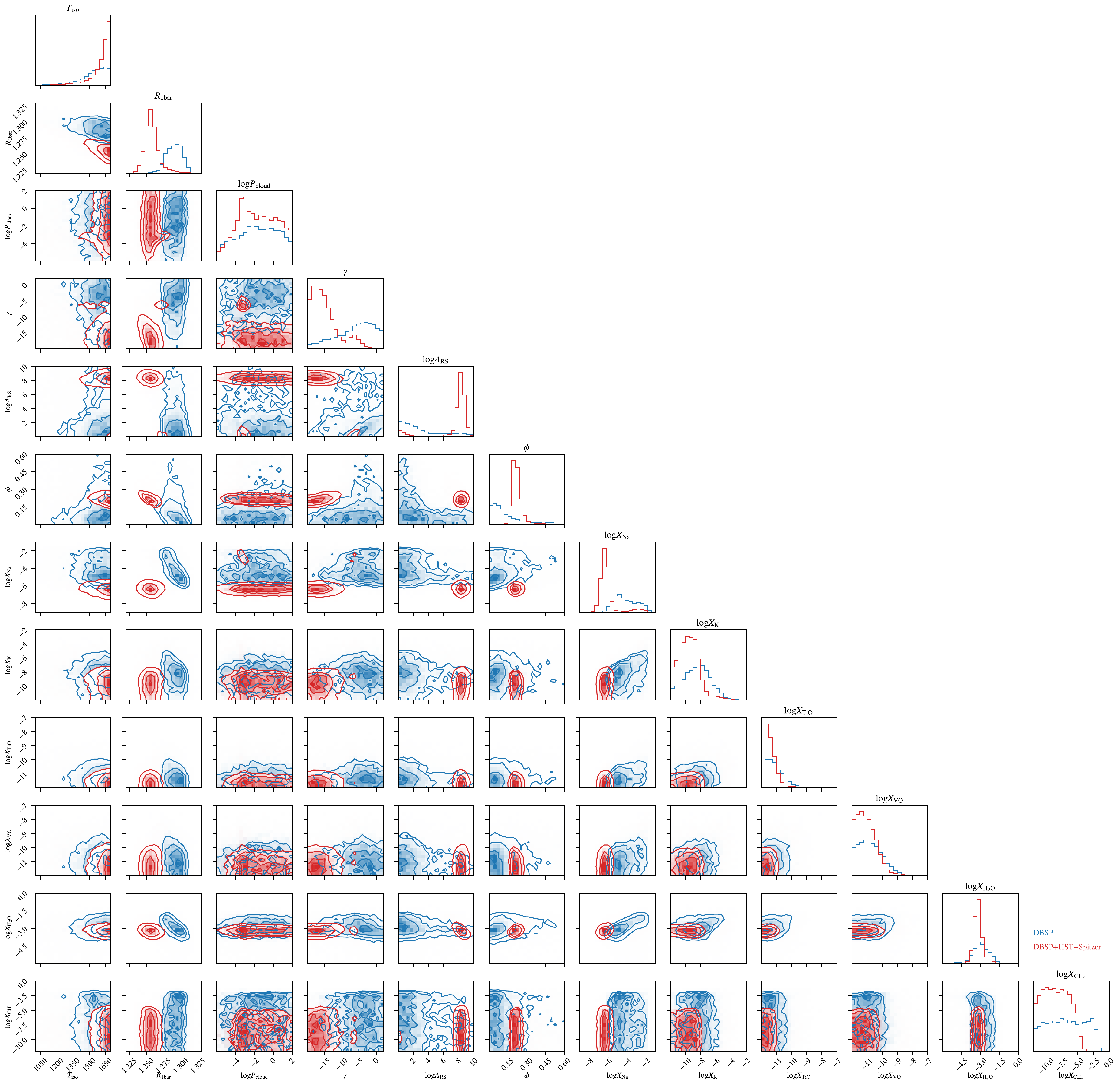}
\caption{Corner plot for spectral retrieval analyses assuming free chemistry.}
\label{fig:corner_freechem}
\end{figure}

The joint posterior distributions of the equilibrium chemistry retrievals are shown in Figure \ref{fig:corner_eqchem}, while those of the free chemistry retrievals in Figure \ref{fig:corner_freechem}. In both figures, the blue contours refer to the retrievals on the DBSP data, while the red ones refer to those on the DBSP, HST, and Spitzer combined data. To save space, the volume mixing ratios of CO, CO$_2$, HCN, and NH$_3$ are omitted in the plot, which are unconstrained.

%++++++++++++++++++++++++++++++++++
%   Figure
%++++++++++++++++++++++++++++++++++

%++++++++++++++++++++++++++++++++++
%   Figure
%++++++++++++++++++++++++++++++++++

%++++++++++++++++++++++++++++++++++
%   Figure
%++++++++++++++++++++++++++++++++++

\bibliography{ref_db.bib}{}
\bibliographystyle{aasjournal}



\end{document}